\documentclass[useAMS,usenatbib,usedcolumn]{mn2e}

\def\HI{\ifmmode{\rm HI}\else{H\/{\sc i}}\fi}

\def\lsun{\ifmmode{{\mathrm L}_{\odot}}\else{L$_{\odot}$}\fi}

\def\msun{\ifmmode{{\mathrm M}_{\odot}}\else{M$_{\odot}$}\fi} 
\def\msunpc2{\ifmmode{{\mathrm M}_{\odot} \, {\mathrm{pc}}^{-2}}\else{M$_{\odot} \, {\mathrm {pc}}^{-2}$}\fi}

\def\kms{\ifmmode{{\mathrm{km \, s^{-1}}}}\else{${\mathrm{km \, s^{-1}}}$}\fi}

\def\la{\mathrel{\mathchoice {\vcenter{\offinterlineskip\halign{\hfil
$\displaystyle##$\hfil\cr<\cr\sim\cr}}}
{\vcenter{\offinterlineskip\halign{\hfil$\textstyle##$\hfil\cr
<\cr\sim\cr}}}
{\vcenter{\offinterlineskip\halign{\hfil$\scriptstyle##$\hfil\cr
<\cr\sim\cr}}}
{\vcenter{\offinterlineskip\halign{\hfil$\scriptscriptstyle##$\hfil\cr
<\cr\sim\cr}}}}}
\def\ga{\mathrel{\mathchoice {\vcenter{\offinterlineskip\halign{\hfil
$\displaystyle##$\hfil\cr>\cr\sim\cr}}}
{\vcenter{\offinterlineskip\halign{\hfil$\textstyle##$\hfil\cr
>\cr\sim\cr}}}
{\vcenter{\offinterlineskip\halign{\hfil$\scriptstyle##$\hfil\cr
>\cr\sim\cr}}}
{\vcenter{\offinterlineskip\halign{\hfil$\scriptscriptstyle##$\hfil\cr
>\cr\sim\cr}}}}}

\usepackage{graphics}
\usepackage{epsfig}
\usepackage{multirow}

\usepackage{bigdelim}
\usepackage{bigstrut}

\usepackage{cite}

\title[Kinematic clues to the origins of S0s]{Planetary Nebula Spectrograph survey of S0 galaxy kinematics. 
II. Clues to the origins of S0 galaxies}
\author[Cortesi et al.]
  {A.~Cortesi$^{1,2,3}$\thanks{email:aricorte@gmail.com}, M.~R.~Merrifield$^1$, L.~Coccato$^2$,
    M.~Arnaboldi$^2$, O.~Gerhard$^3$, 
      \newauthor  
   S.~Bamford$^1$, N.~R.~Napolitano$^4$,  A.~J.~Romanowsky$^5$, N.~G.~Douglas$^6$,
  \newauthor  
  K.~Kuijken$^7$,  M.~Capaccioli$^8$, K.~C.~Freeman$^9$, K.~Saha$^3$,  and A.~L. Chies-Santos$^1$\\ 
   $^1$University of Nottingham, School of Physics and Astronomy, University
       Park, NG7 2RD Nottingham, UK \\
   $^2$European Southern Observatory, Karl-Schwarzschild-Strasse 2, 85748
       Garching, Germany \\ 
   $^3$Max-Planck-Institut f\"ur Extraterrestrische Physik,
       Giessenbachstrasse,  85741 Garching, Germany \\
   $^4$Istituto Nazionale di Astrofisica, Osservatorio Astronomico di
       Capodimonte, Via Moiariello 16, 80131 Naples, Italy \\
   $^5$Department of Physics and Astronomy, San Jos\'e State University,
	One Washington Square, San Jose, CA, 95192, USA \\
	and University of California Observatories, 1156 High St., Santa Cruz, CA 95064, USA\\
   $^6$Kapteyn Astronomical Institute, University of Groningen, PO Box 800,
       9700 AV Groningen, The Netherlands \\
   $^7$Leiden Observatory, Leiden University, PO Box 9513, 2300 RA Leiden, The
       Netherlands \\
   $^8$Dipartimento di Fisica, Universit\`a ``Federico II'', Naples, Italy \\
   $^9$Research School of Astronomy and Astrophysics, Australian National
       University, Canberra, Australia} 

\begin{document}

\date{; }

\maketitle

\begin{abstract}
 
  The stellar kinematics of the spheroids and discs of S0 galaxies
  contain clues to their formation histories. Unfortunately, it is
  difficult to disentangle the two components and to recover their
  stellar kinematics in the faint outer parts of the galaxies using
  conventional absorption line spectroscopy. This paper therefore
  presents the stellar kinematics of six S0 galaxies derived from
  observations of planetary nebulae (PNe), obtained using the
  Planetary Nebula Spectrograph.  To separate the kinematics of the
  two components, we use a maximum-likelihood method that combines the
  discrete kinematic data with a photometric component decomposition.
  The results of this analysis reveal that: the discs of S0 galaxies
  are rotationally supported; however, the amount of random motion in
  these discs is systematically higher than in comparable spiral
  galaxies; and the S0s lie around one magnitude below the
  Tully--Fisher relation for spiral galaxies, while their spheroids
  lie nearly one magnitude above the Faber--Jackson relation for
  ellipticals.  All of these findings are consistent with a scenario
  in which spirals are converted into S0s through a process of mild
  harassment or ``pestering,'' with their discs somewhat heated and
  their spheroid somewhat enhanced by the conversion process.
  In such a scenario, one might expect the properties of S0s to depend on environment.  We do not see such an effect in this fairly small sample, although any differences would be diluted by the fact that the current location does not necessarily reflect the environment in which the transformation occurred.  Similar observations of larger samples probing a broader range of environments, coupled with more detailed modelling of the transformation process to match the wide range of parameters that we have shown can now be measured, should take us from these first steps to the definitive answer as to how S0 galaxies form. 
\end{abstract}

\begin{keywords}
  galaxies: elliptical and lenticular -- galaxies: evolution --
  galaxies: kinematics and dynamics.
  \vspace{2 cm}
\end{keywords}

\vspace{2 cm}

\section{Introduction}
\label{sec:introduction}

The origin of lenticular, or S0, galaxies, and in particular whether they are
more closely related to spiral galaxies or elliptical galaxies,
remains obscure.  They display the bulge-plus-disc morphology that we
associate with spiral galaxies, but they lack the young stars in
spiral arms that we associate with such systems, and tend to be more
dominated by their spheroidal bulge components, making it tempting to
associate them with elliptical galaxies.  

\begin{figure}
\includegraphics[width=0.45\textwidth]{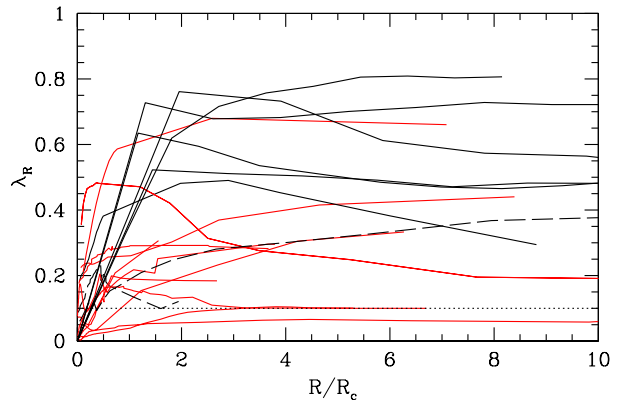}\\
\caption{Plot showing a measure of specific angular momentum as a
  function of radius for a sample of early-type galaxies.  Black solid
  lines are S0s from \citet{Ari}, black dashed lines are S0s from
  \citet{Coccato}, and red lines are ellipticals from \citet{Coccato}.
  The dotted line shows the suggested separation between slow and fast
  rotators as a kinematic classification scheme
  \citep{Emsellem2007}.\label{fig:lambda}}
\end{figure}

Moving away from purely morphological considerations, we might hope
that kinematic information can help to ascertain which other galaxies
are S0s closest cousins, since the traditional view is that
disc-dominated spiral galaxies will reflect that morphology in
rotationally-dominated kinematics, while elliptical galaxies are
largely supported by random motions.  However, more recent data
suggests that the situation is somewhat subtler.  The importance of
rotation was quantified by \citet{Emsellem2007}, who defined the
quantity $\lambda_{R}$, the cumulative specific angular momentum of
galaxies as a function of radius, seeking to draw a distinction
between disc- and bulge-dominated systems via this
parameter. Unfortunately, they show that there is no clear dichotomy
in this quantity, with a broad spectrum of degrees of rotational
support.  \citet{Coccato} calculate the same quantity for a sample of
elliptical galaxies, but using PNe as tracers of the velocity field
and, in this way, exploring a wider range of radii. They show that the
recovered $\lambda_{R}$ smoothly join the quantity obtained with the
SAURON data, which describe the central regions of the galaxies, where
PNe are not detectable. This radially-extended reconstruction of the
specific angular momentum for a sample of elliptical \citep{Coccato}
and S0 \citep{Ari} galaxies, shown in Figure~\ref{fig:lambda},
displays the same continuous range of profiles, seemingly unrelated
to the galaxy morphology.  It also presents a surprising degree of
variation in profile shape at the large radii probed by PNe, with some
dropping dramatically, apparently changing from
rotationally-supported systems to being dominated by random motions in
their outermost parts.  

We first encountered this phenomenon in an analysis of the S0 galaxy
NGC~1023, where a pilot study of PNe with the Planetary Nebula
Spectrograph (PN.S) showed a major decrease in rotational motion at
large radii \citep{EDO}.  This drop in angular momentum was initially
interpreted as evidence that a normal rotationally-supported stellar
disc had been dramatically heated into random motion by a merger in
its outer parts.  However, by more careful modelling of separate bulge
and disc components, we were able to show that this behaviour was in
fact caused simply by the superposition of a cold rotating disc and a
hot spheroidal component, whose relative contributions to the observed
kinematics varied with both radius and azimuth around the galaxy
\citep{Ari}.

This discovery underlined the importance of treating S0 galaxies as
multi-component systems (see also \citet{Kormedy2012}), and taking care to separate out the distinct
stellar bulge and disc elements kinematically as well as
photometrically.  Once this decomposition has been performed properly,
we have access to a range of diagnostics that can be used to provide
clues to the formation history of these galaxies.  For example, if
they are simply spiral galaxies that have quietly ceased forming
stars, one would expect their stellar discs to have the same kinematic
properties as those in spirals; if, on the other hand, a more violent
event such as a minor merger led to the transition from spiral to S0,
one might expect the disc, if not destroyed in the process, to at
least have had its random motions increased significantly
\citep{Bournaud}.  Equally, with reliable kinematic parameters for the
individual components, we can see where they lie relative to the usual
kinematic scaling relations followed by spirals and ellipticals, to
see if they could plausibly have evolved from such progenitors.

In this paper, we therefore apply the kinematic bulge--disc
decomposition technique developed in \citet{Ari} to the sample of six
S0s with suitable PNe data presented in \citet{Ari2}.  This sample was
selected to span a range of environments in order to see if the
evolutionary properties depend systematically on surroundings:
NGC~7457 and NGC~3115 are isolated; NGC~1023 and NGC~2768 are the
dominant galaxies of two small groups (5-6 members); while NGC~3384
and NGC~3489 are satellite galaxies of the Leo Group (30 members).
The Hubble types of these galaxies and the adopted distance moduli are
listed in Table~\ref{tab:galfitinput}.

The remainder of this paper is laid out as follows.  In
Section~\ref{sec:decomposition}, we present the photometric
bulge--disc decomposition that underlies the analysis, and confirms
the nature of these galaxies as composite S0 systems.
Section~\ref{sec:disc and spheroid kinematics} then uses this
decomposition to model the spheroid and disc kinematics of each
system, and Section~\ref{sec:study of the recovered kinematics} looks
at the characteristic kinematics of the separate components for
indications as to their origins.  Section~\ref{sec:Discussion} brings
this evidence together to present the emerging picture as to the steps
leading to the formation of a lenticular galaxy.

\section{Photometric decomposition }
\label{sec:decomposition} 

In order to correctly assign each PN's observed velocity to disc or
bulge kinematics, we must first calculate the relative contributions
of disc and bulge to the total light at its location.  To do so, we
follow the same photometric decomposition procedure described in
\citet{Ari}, briefly summarised here for completeness.

\begin{table*}
\begin{footnotesize}
\centering
\begin{tabular}{c|cccccc}
\hline
Name & Type  & Distance modulus &  Archive  & Band  & Angular scale    & Zero point \\
$$ & $$ & [mag] &  $$     &$$ & ['' per pixel] & [mag] \\
\hline
\hline
NGC~3115  &  S0-edge-on & $29.77$ & 2MASS &$K$&$1$ & $20.86$ \\
NGC~7457  &  SA(rs)0$^{-}$ &  $30.45$ & 2MASS &$K$&$1$ & $20.09$ \\
NGC~2768  &  E6 &  $31.59$ & 2MASS &$K$ & $1$&$20.10$ \\
NGC~1023  &  SB(rs)0$^{-}$ &    $30.13$ & 2MASS &$K$&$1$ & $20.03$ \\
NGC~3489  &  SAB(rs)0$^{+}$ &  $30.25$ & SDSS &$z$&$0.4$ & $22.55$ \\
NGC~3384  &  SB(s)0$^{-}$ &  $30.16$ & 2MASS &$K$&$1$ & $20.09$ \\
\hline
\end{tabular}
\caption{Basic data on the sample and images analysed. From left to
  right: galaxy name, galaxy type, distance modulus from \citet{Tonry}
  shifted by 0.16 magnitudes (see text for details), archive, band,
  angular scale, zero point \label{tab:galfitinput}}
\end{footnotesize}
\end{table*}

Where possible, the galaxies' images used for the decomposition are
$K$-band data from the 2MASS survey \citep{Skrutskie+06},
since these near infrared images minimize the influence of any modest
amounts of dust extinction present in these S0 systems.  For NGC~3489,
the 2MASS image did not go deep enough to carry out the decomposition,
so we instead used the reddest SDSS optical image, and for NGC~1023 we
also employed a deep R-band image \citep{EDO}, which allowed us to
mask the companion galaxy NGC~1023A while performing the decomposition
\citep{Ari}.  The details of the images and other input data used in
the fitting process are presented in Table~\ref{tab:galfitinput}.

\begin{table*}
\begin{footnotesize}
\centering
\begin{tabular}{c|ccccc|ccccc|c}
\hline
\multicolumn{6}{c}{Disc} & \multicolumn{5}{c}{Spheroid} & \multicolumn{1}{c}{B/T}\\
\hline
Name & $m_{D}$  & $R_{d}$ &  $b/a$  &$incl$  & $PA$    & $m_{B}$  & $R_{e}$  & $n$ & $b/a$ & $PA$ & $ $\\
$$ & [mag] & [arcs] &  $$     &[deg] & [deg] & [mag] & [arcs] & $$  & $$    &[deg] & $ $\\
\hline
\hline
NGC~3115  &  $8.34$ & $53.69$ & $0.39$ &$67$ & $45.00$  & $7.17$  & $26.19$ & $4$    & $0.31$ & $45.00$ & $0.74$ \\
NGC~7457  &  $8.56$ & $27.07$ & $0.48$ &$62$ & $-57.28$ & $9.49$  & $11.62$ & $4$    & $0.62$ & $-46.04$ & $0.30$\\ 
NGC~2768  &  $8.19$ & $42.93$ & $0.29$ &$73$ & $-86.25$ &  $7.23$ & $50.46$ & $4.65$ & $0.66$ & $-85.39$ & $0.71$\\ 
NGC~1023  &  $7.02$ & $59.08$ & $0.26$ &$74$ & $84.12$  &  $6.9$  & $17.86$ & $4$    & $0.75$ & $75.59$  & $0.53$\\    
NGC~3489  &  $8.18$ & $24.98$ & $0.49$ &$61$ & $-15.05$ &  $8.05$ & $7.64$ & $4$    & $0.72$ & $-19.50$ & $0.54$\\
NGC~3384  &  $7.92$ & $63.73$ & $0.34$ &$70$ & $52.50$  &  $7.29$ & $15.20$ & $4$    & $0.83$ & $60.51$ & $0.64$  \\
\hline
\end{tabular}
\caption{Results from GALFIT fit. From left to right: galaxy name [1],
  disc apparent magnitude [2], disc scale length [3], disc axes ratio
  [4], used to obtain the galaxy inclination [5], disc position angle
  [6], spheroid apparent magnitude [7], effective radius [8], Sersic
  index, equal to $4$ where this value fitted well, [9], spheroid axes
  ratio [10], spheroid position angle [11] and bulge-to-total light
  ratio [12].  For NGC~3489 the recovered magnitudes in z-band have
  been colour corrected.\label{tab:galfit}}
\end{footnotesize}
\end{table*}

 \begin{figure*}
\begin{tabular}{lcr} 
\includegraphics[width=0.2\textwidth,height=0.2\textwidth]{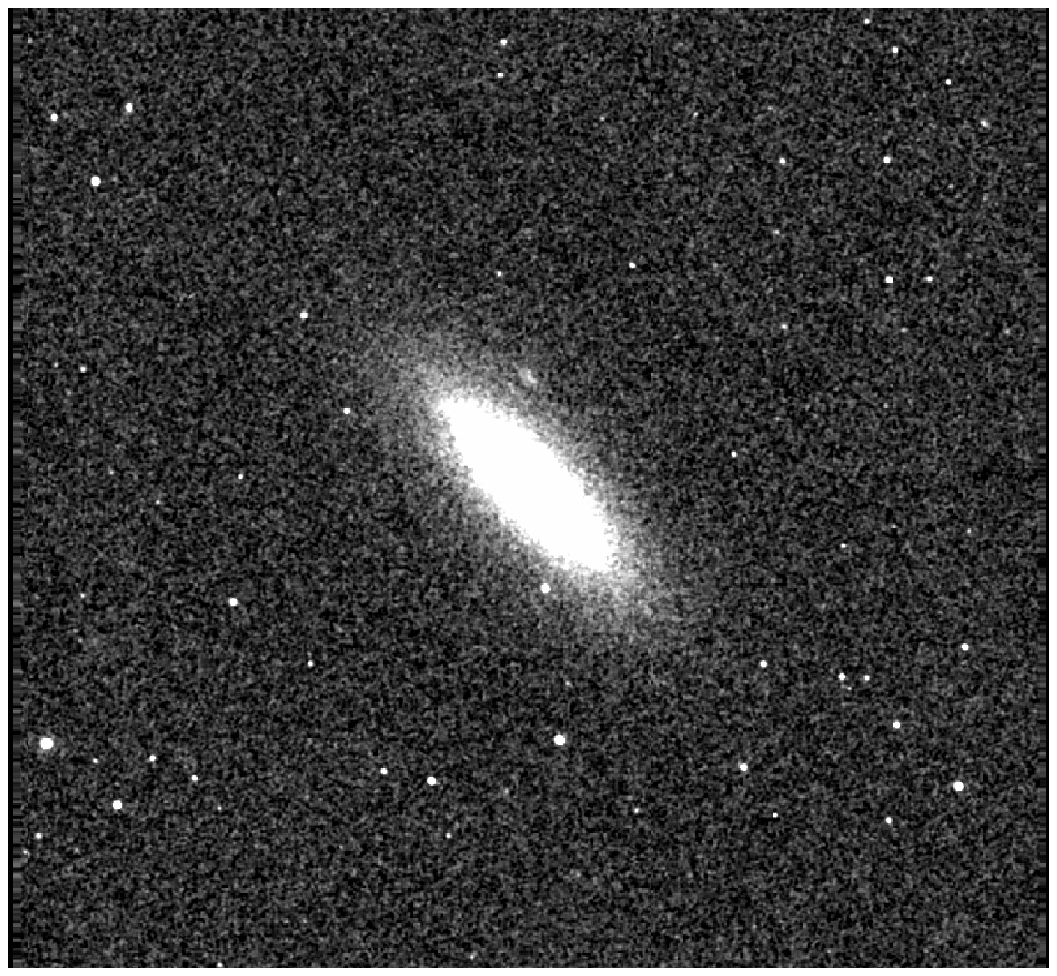}
\includegraphics[width=0.2\textwidth,height=0.2\textwidth]{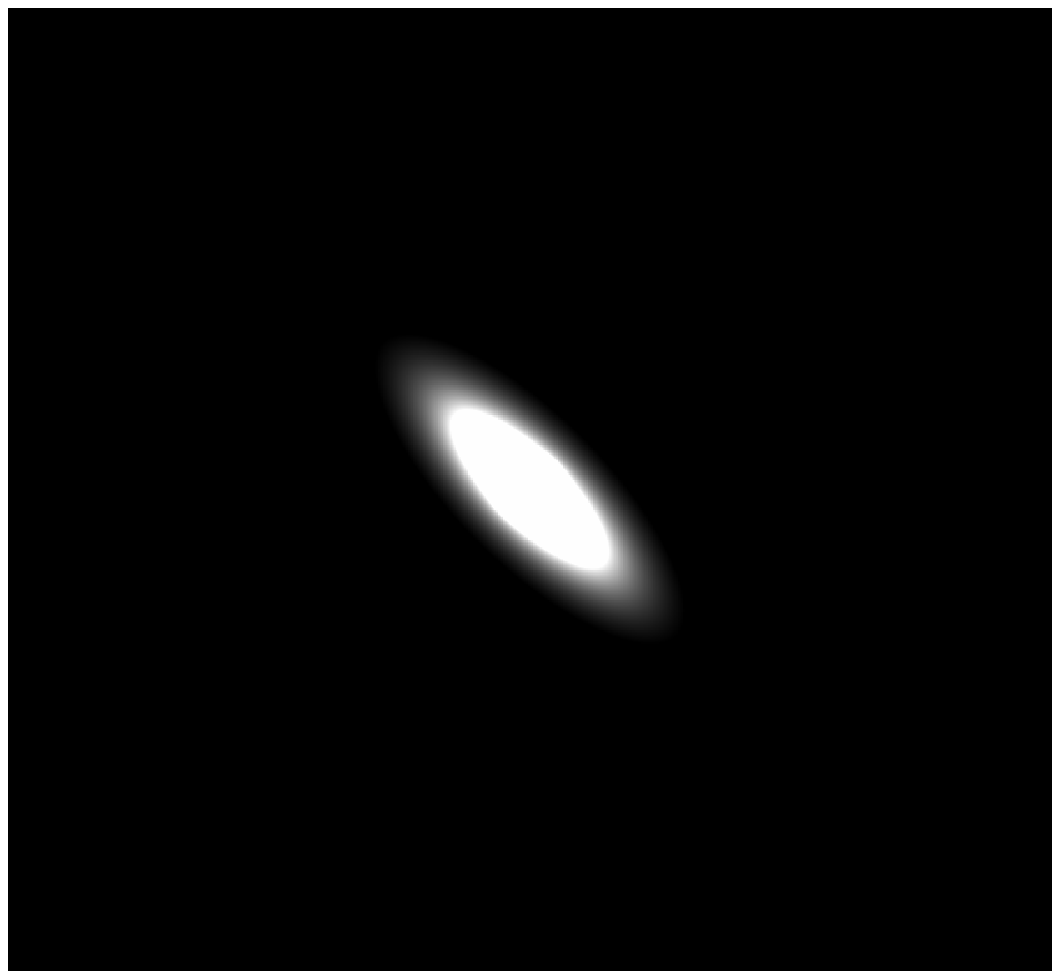}
\includegraphics[width=0.2\textwidth,height=0.2\textwidth]{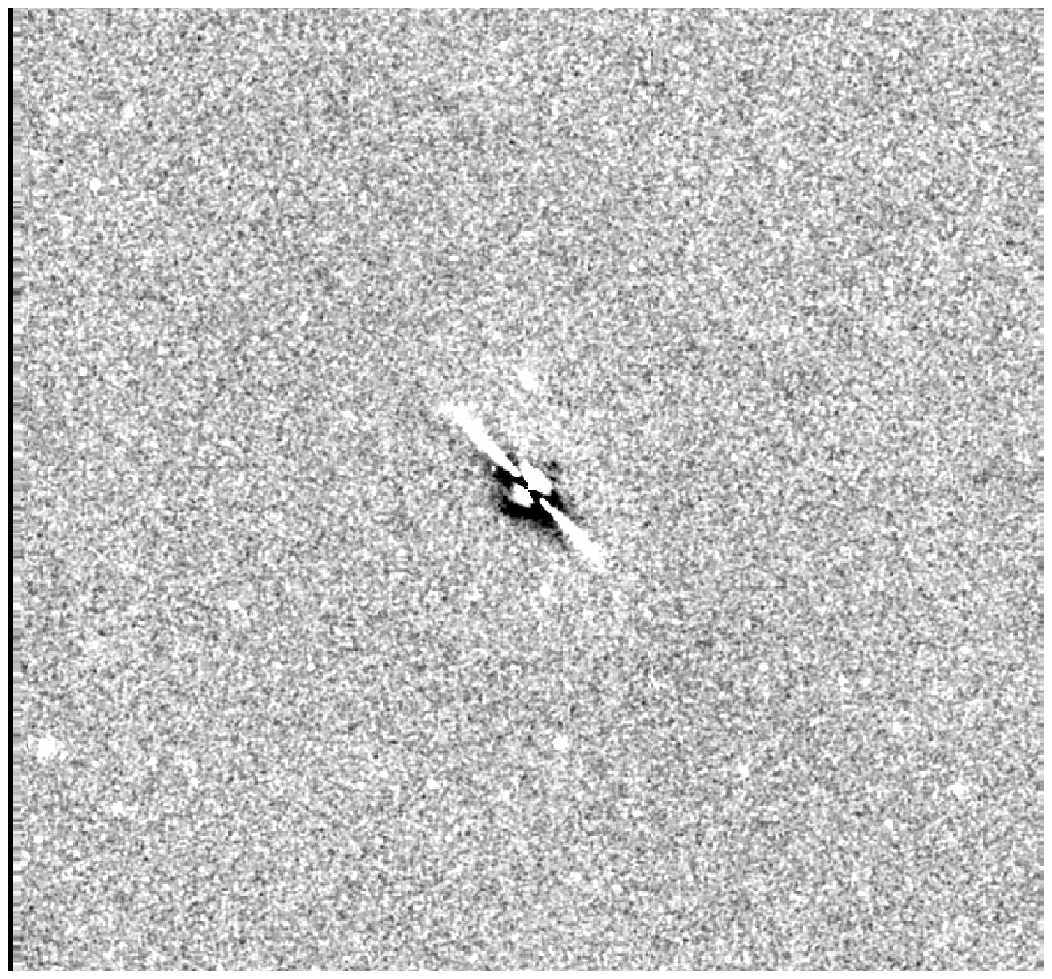}\\
\includegraphics[width=0.2\textwidth,height=0.2\textwidth]{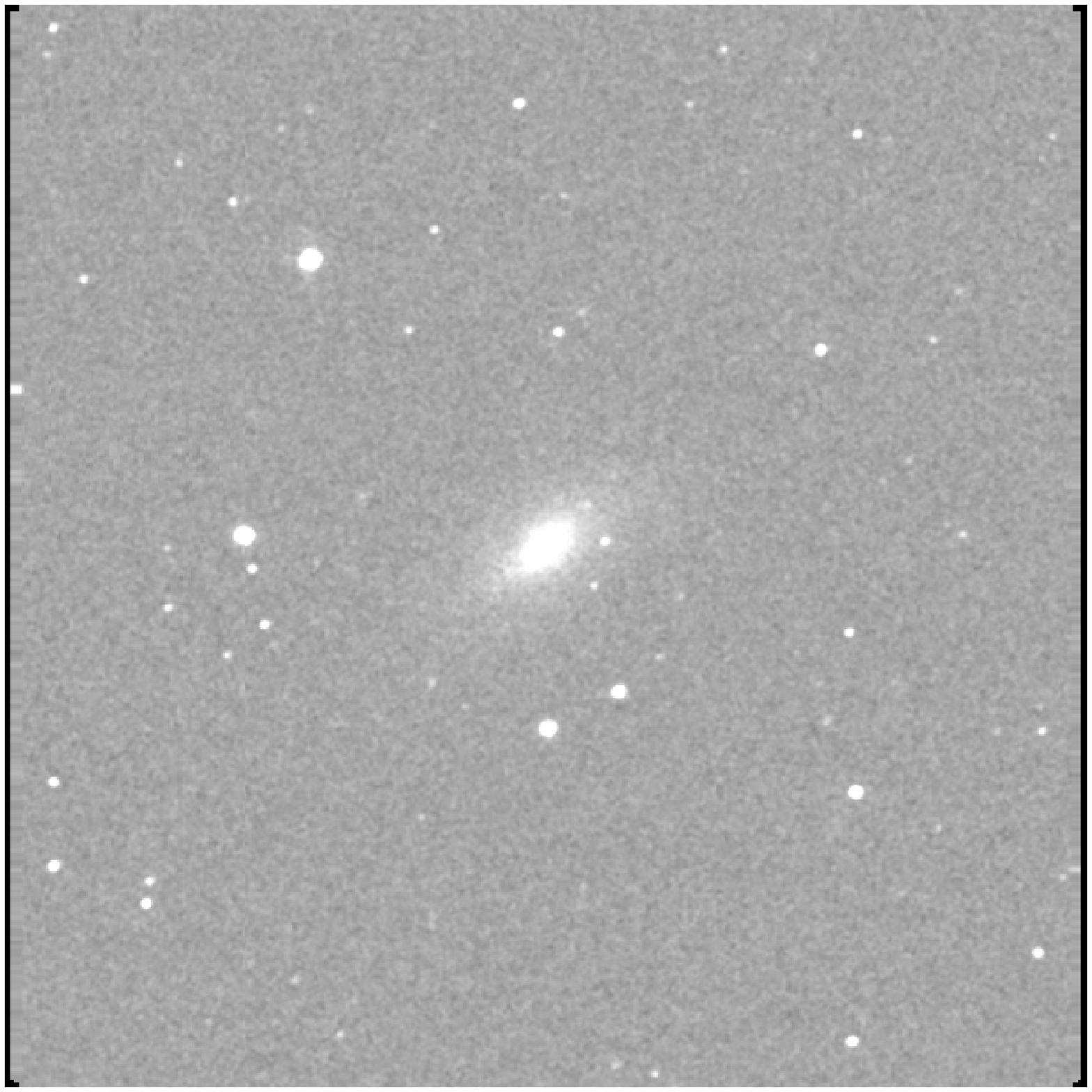}
\includegraphics[width=0.2\textwidth,height=0.2\textwidth]{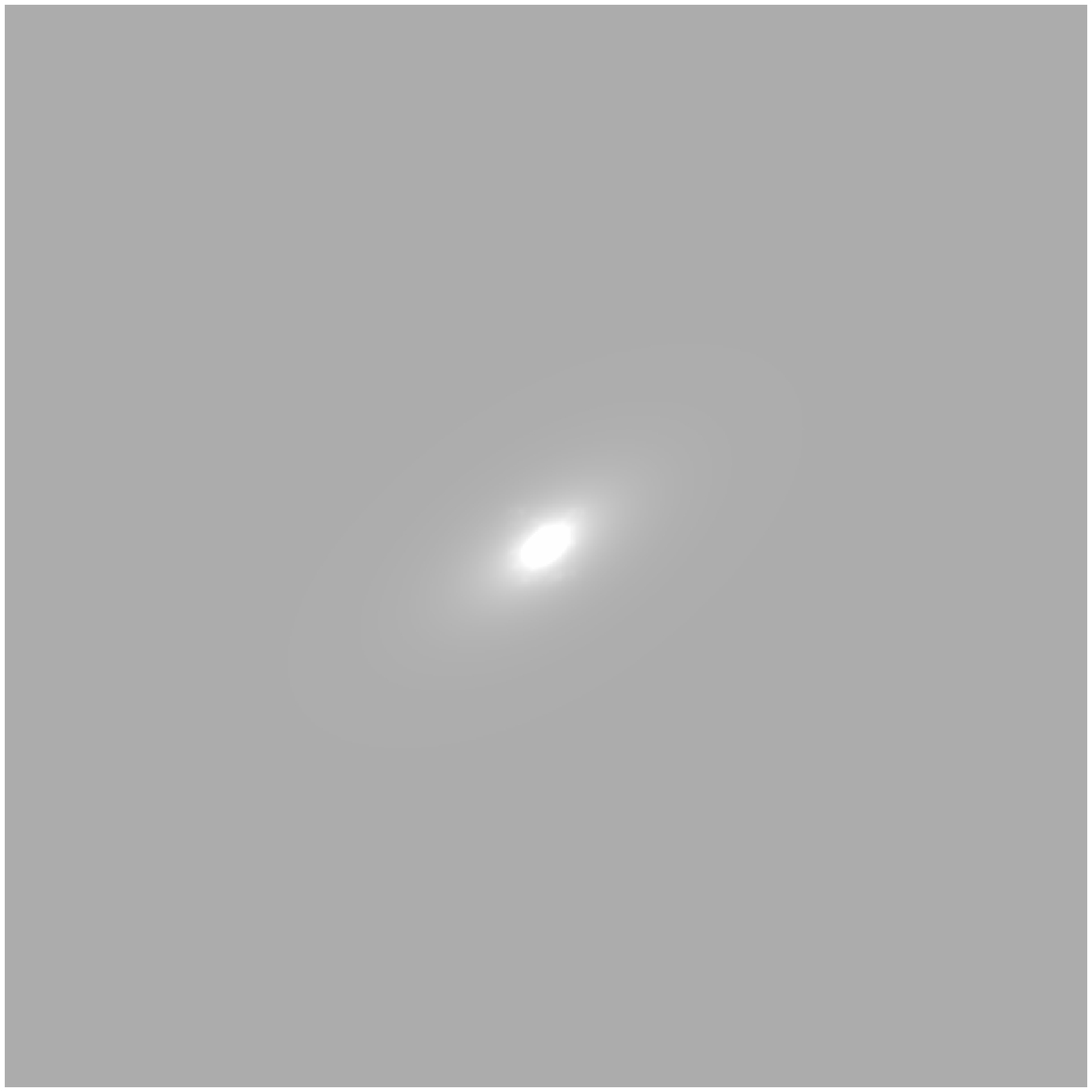}
\includegraphics[width=0.2\textwidth,height=0.2\textwidth]{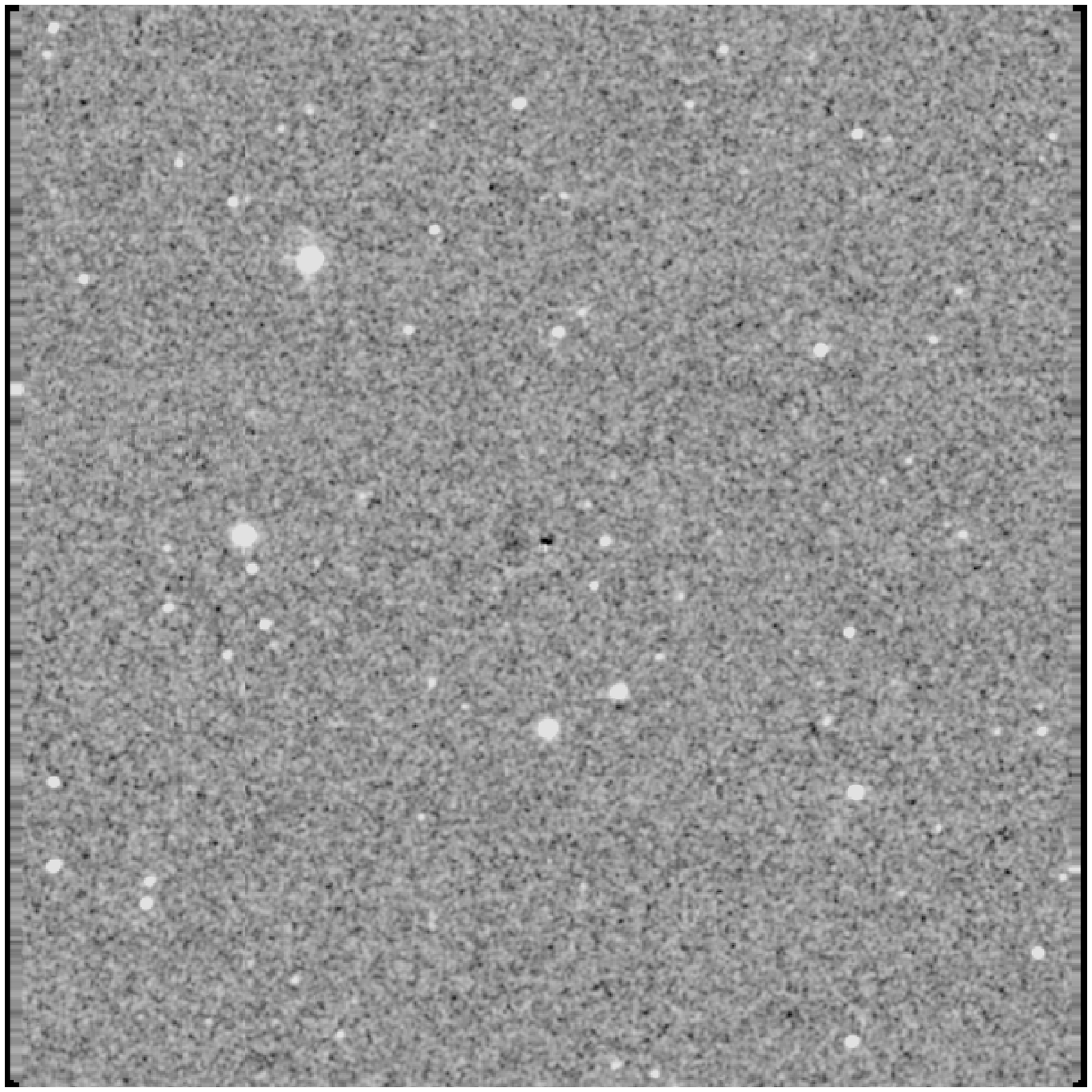}\\
\includegraphics[width=0.2\textwidth,height=0.2\textwidth]{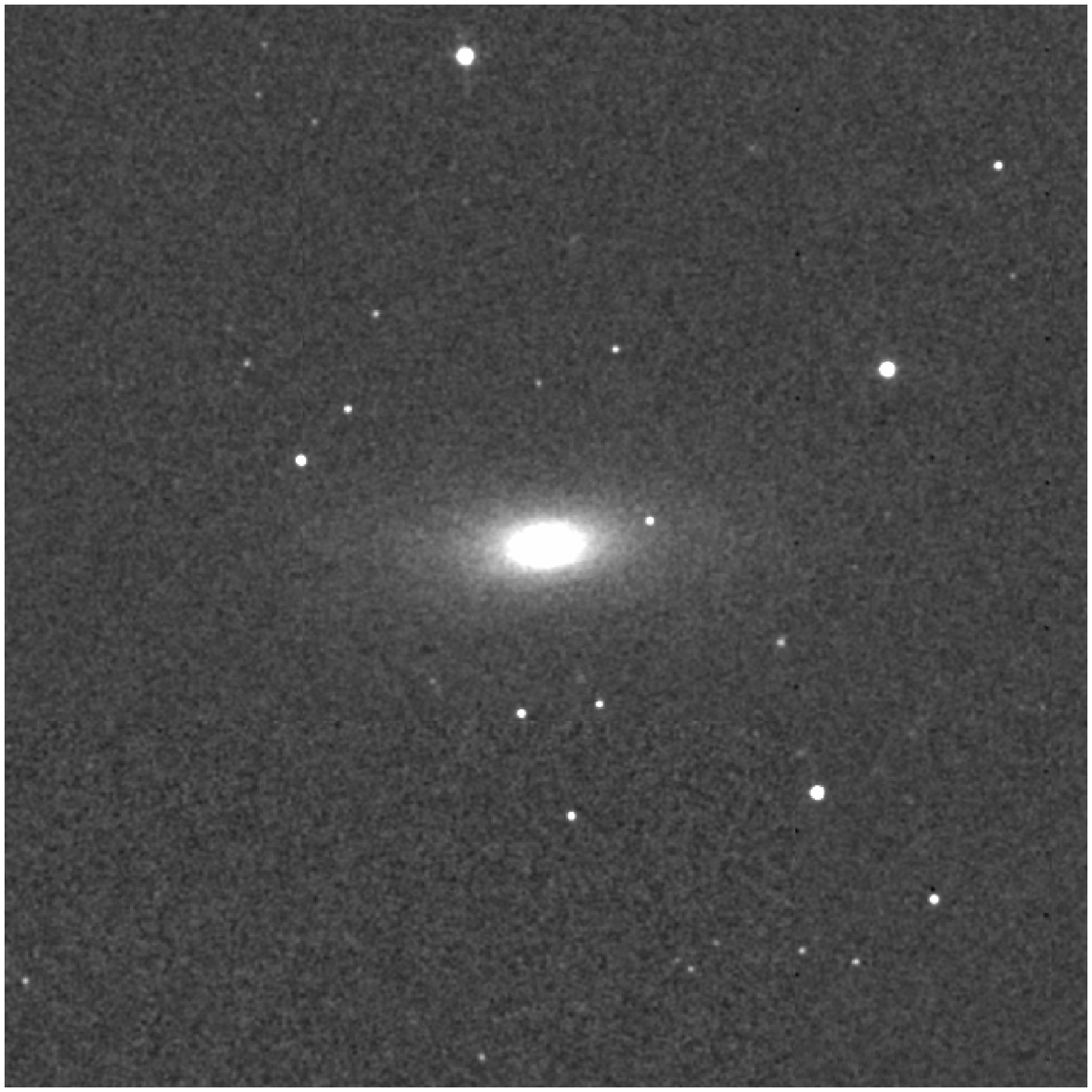}
\includegraphics[width=0.2\textwidth,height=0.2\textwidth]{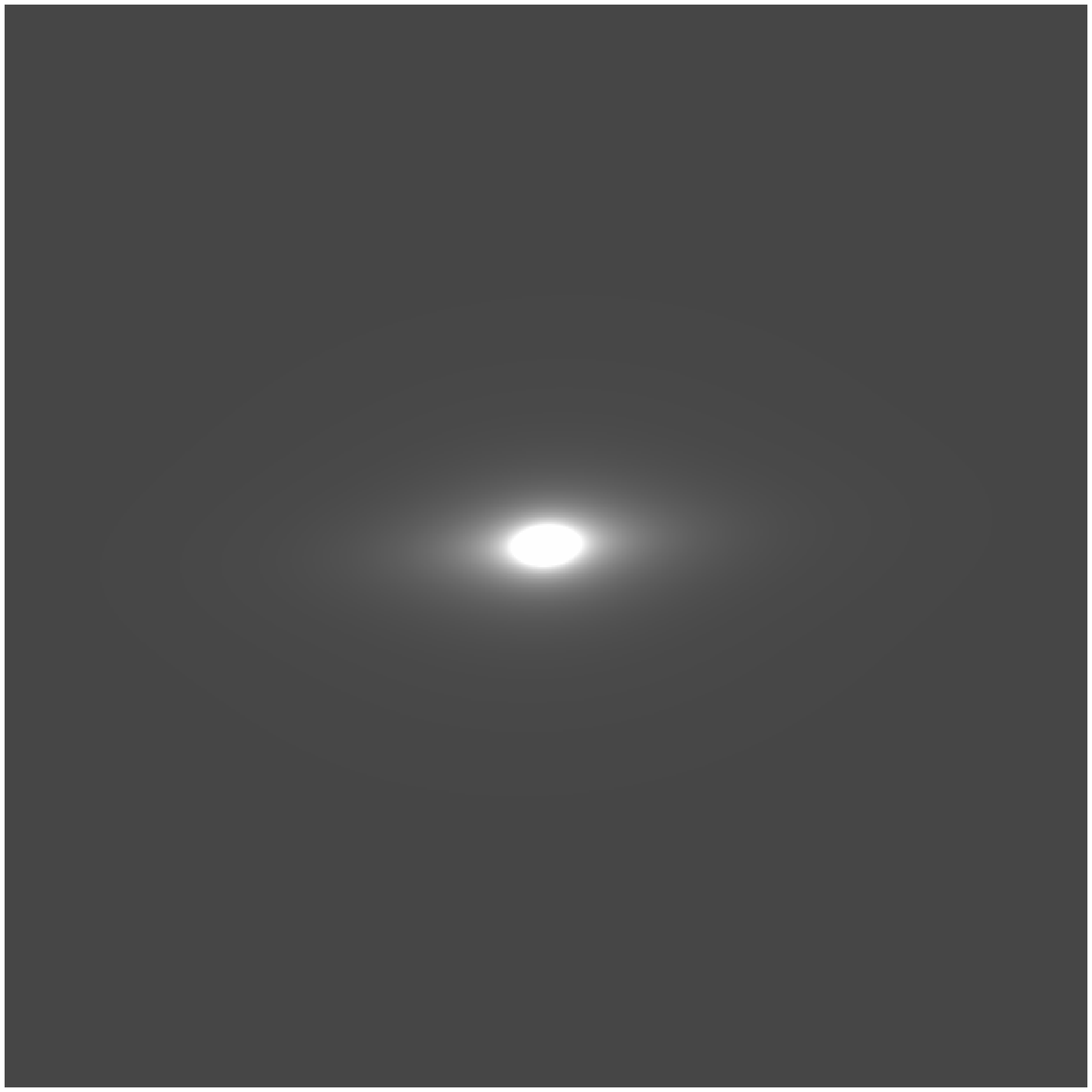}
\includegraphics[width=0.2\textwidth,height=0.2\textwidth]{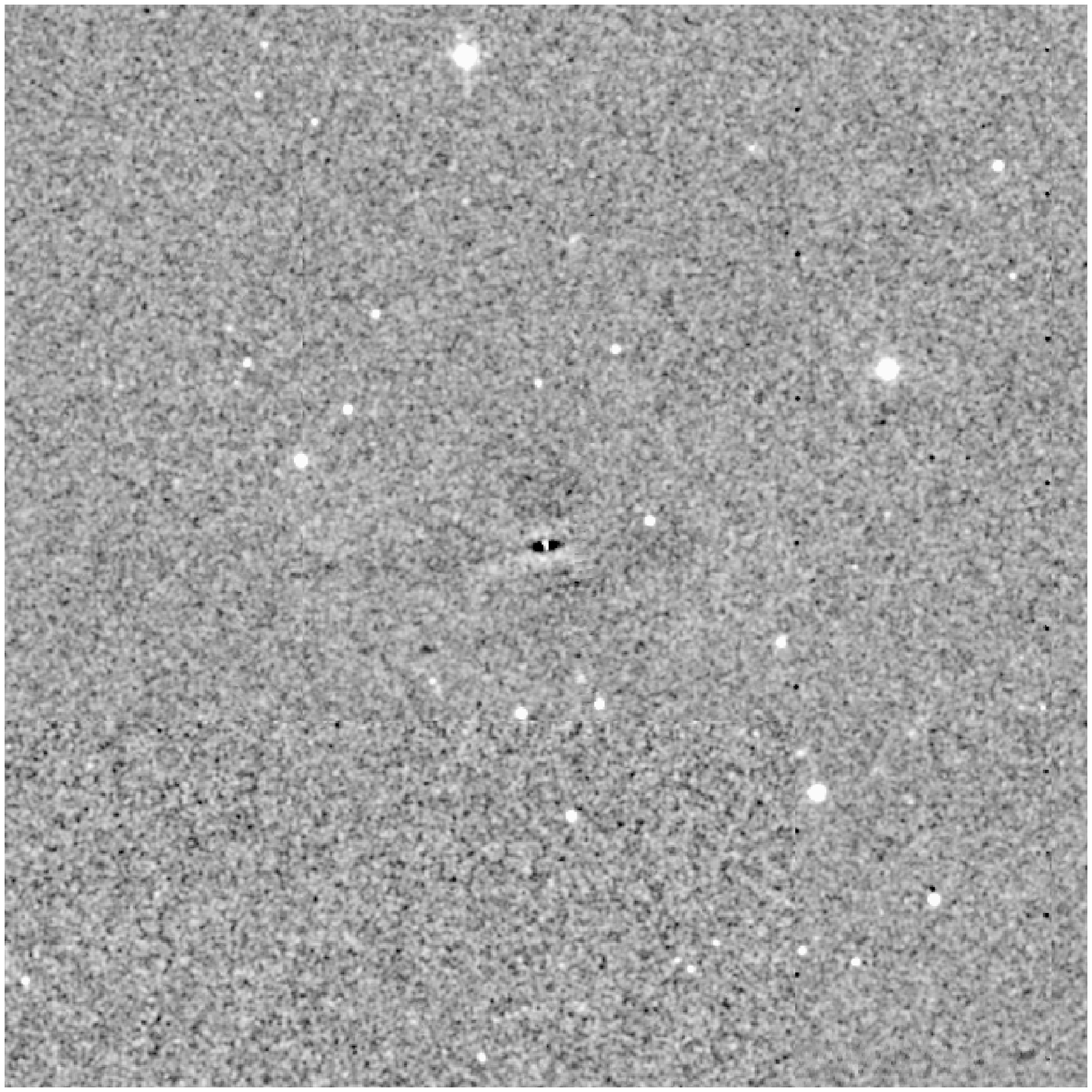}\\
\includegraphics[width=0.2\textwidth,height=0.2\textwidth]{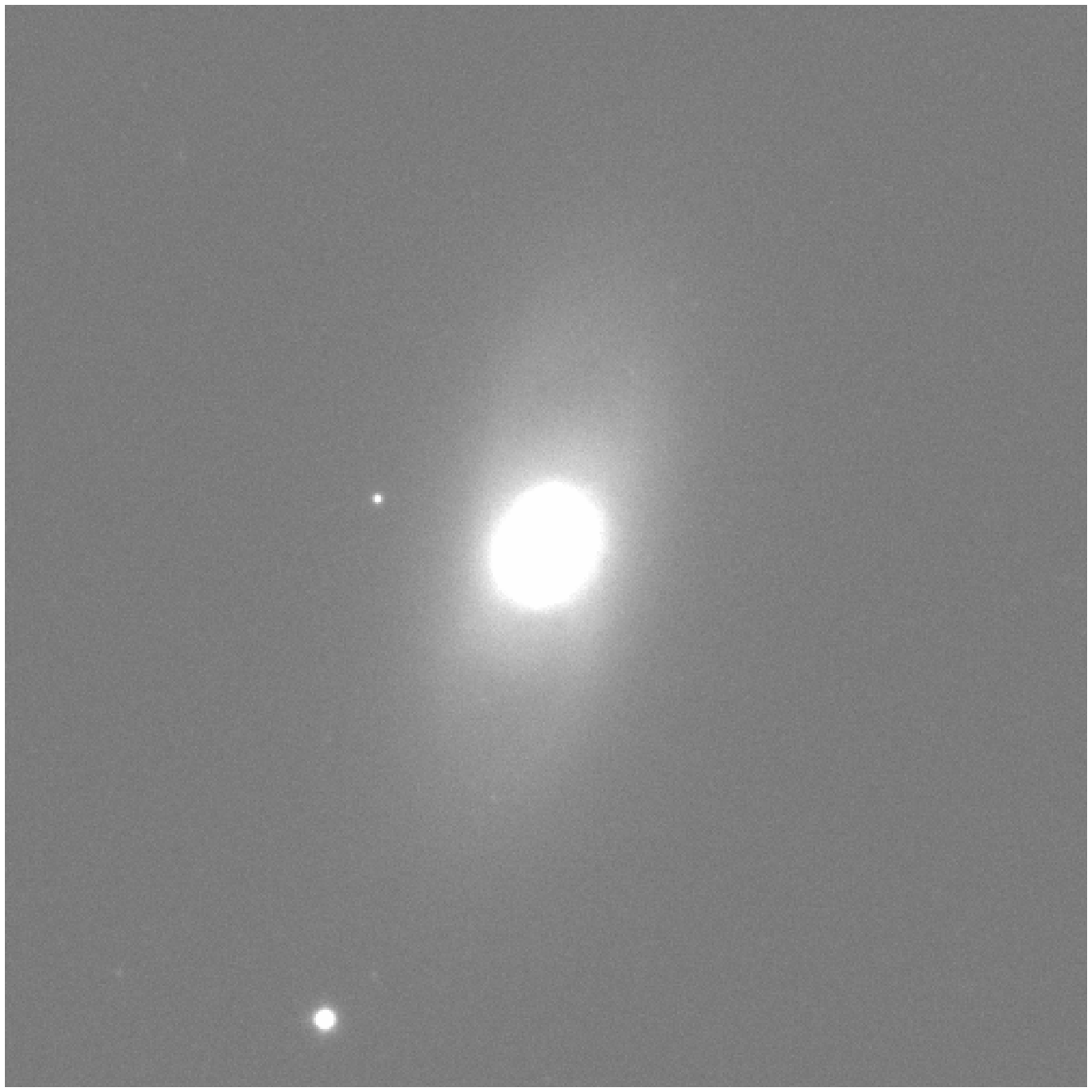}
\includegraphics[width=0.2\textwidth,height=0.2\textwidth]{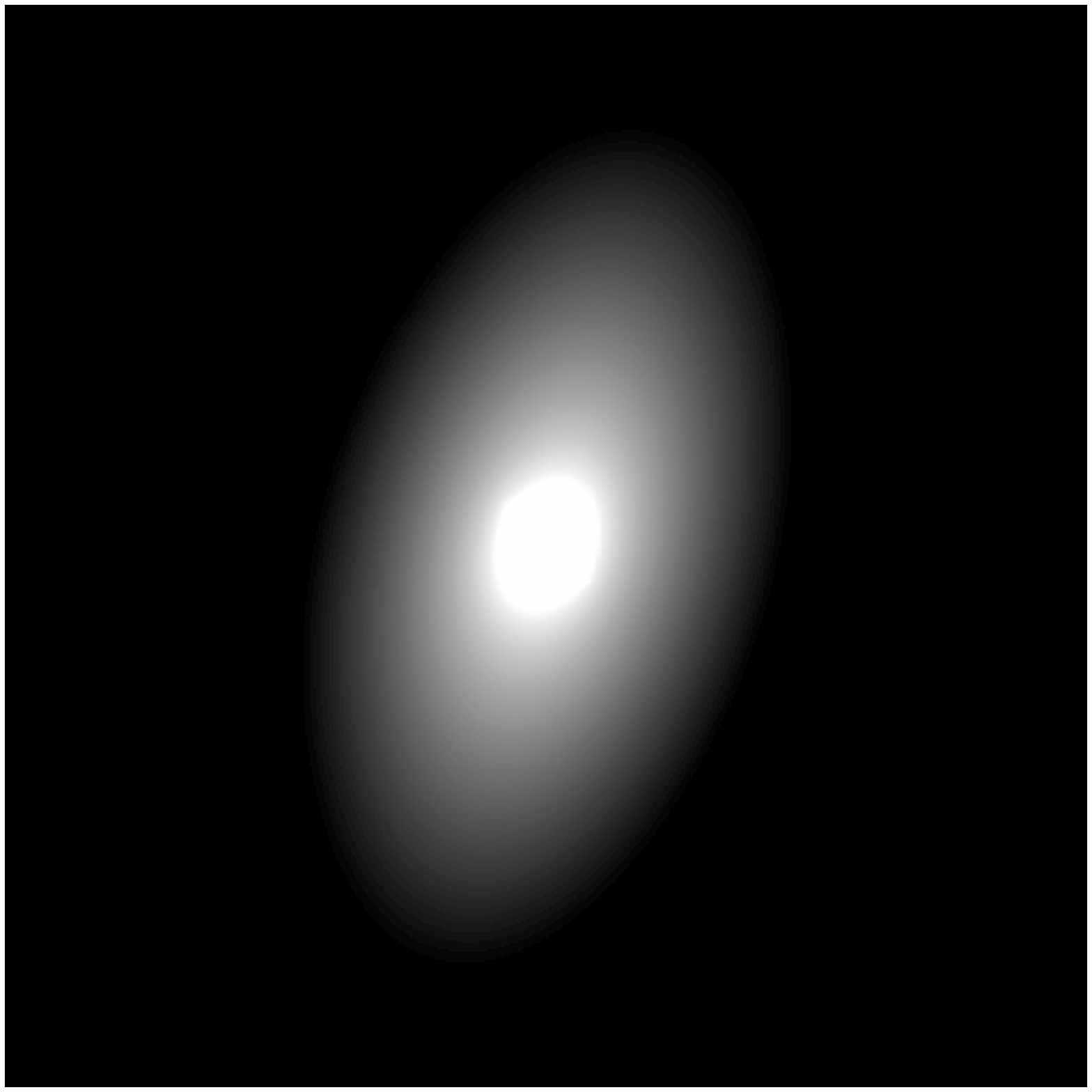}
\includegraphics[width=0.2\textwidth,height=0.2\textwidth]{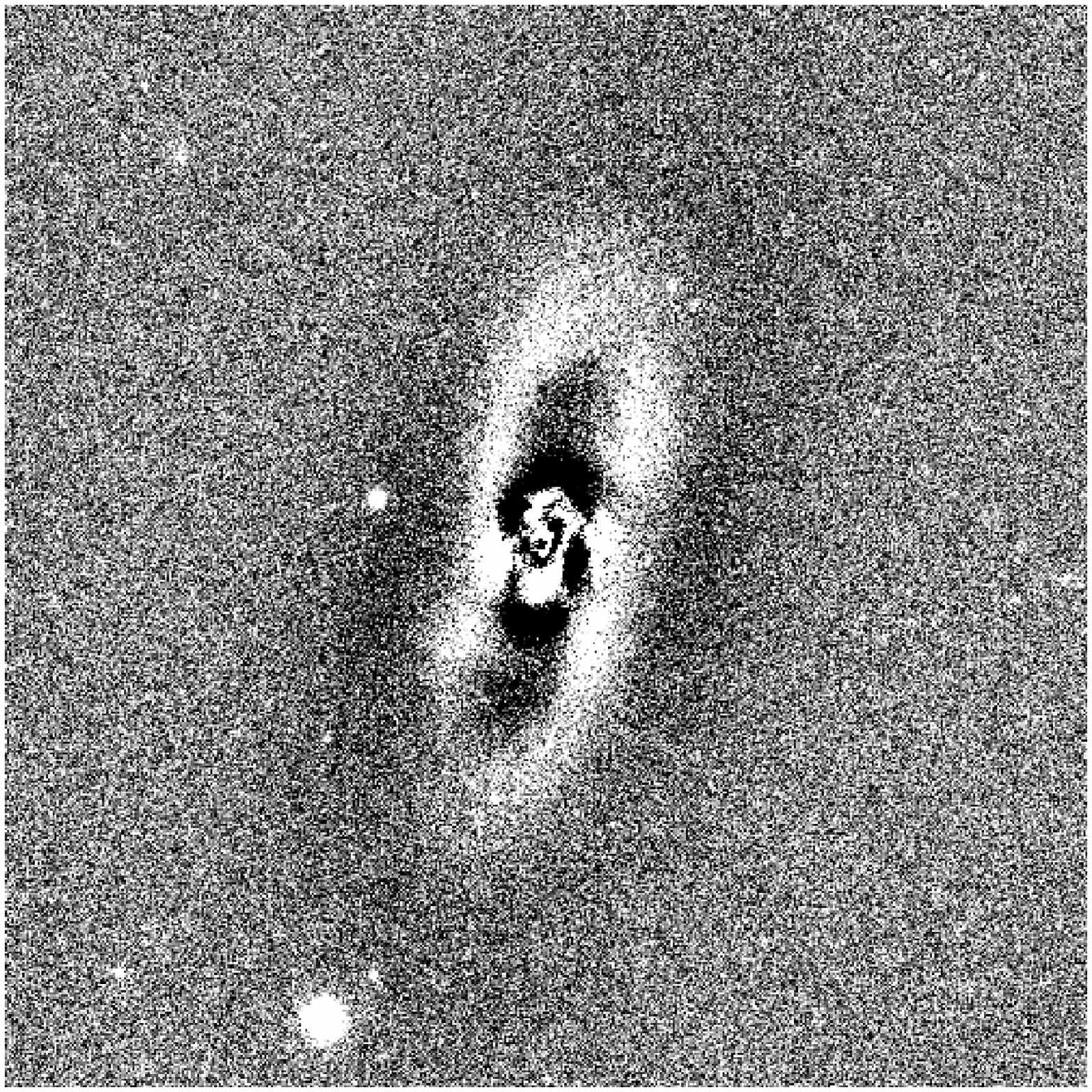}\\
\includegraphics[width=0.2\textwidth,height=0.2\textwidth]{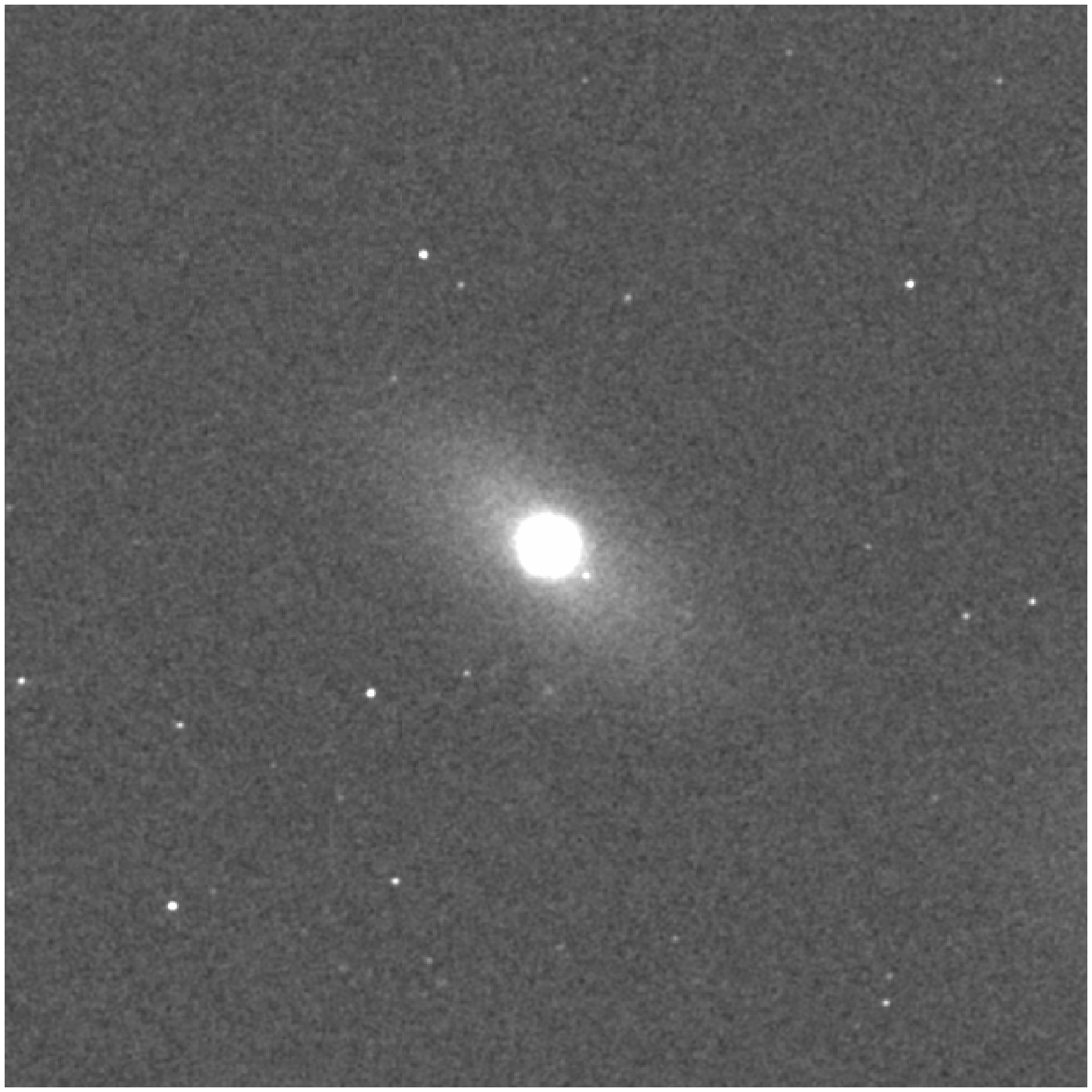}
\includegraphics[width=0.2\textwidth,height=0.2\textwidth]{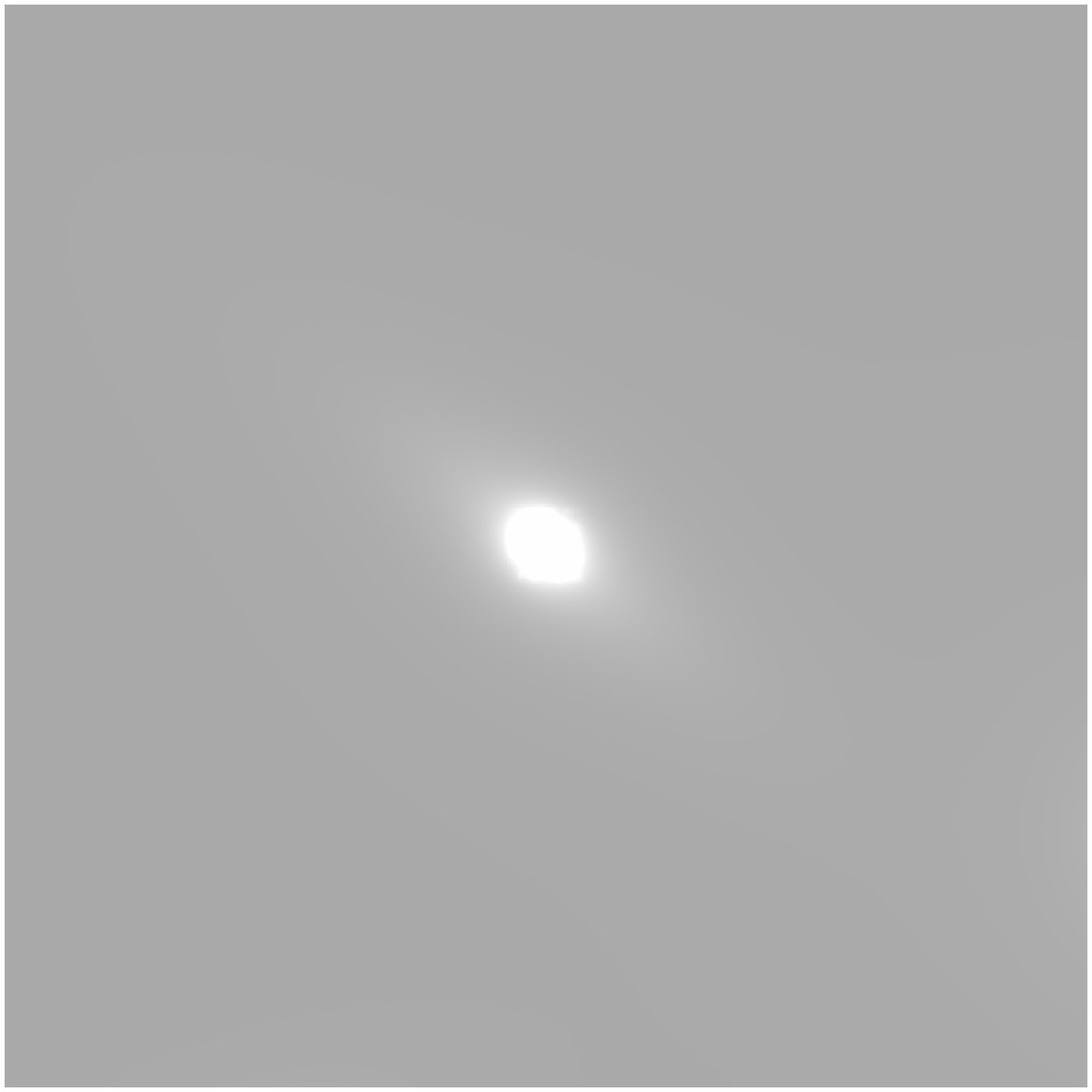}
\includegraphics[width=0.2\textwidth,height=0.2\textwidth]{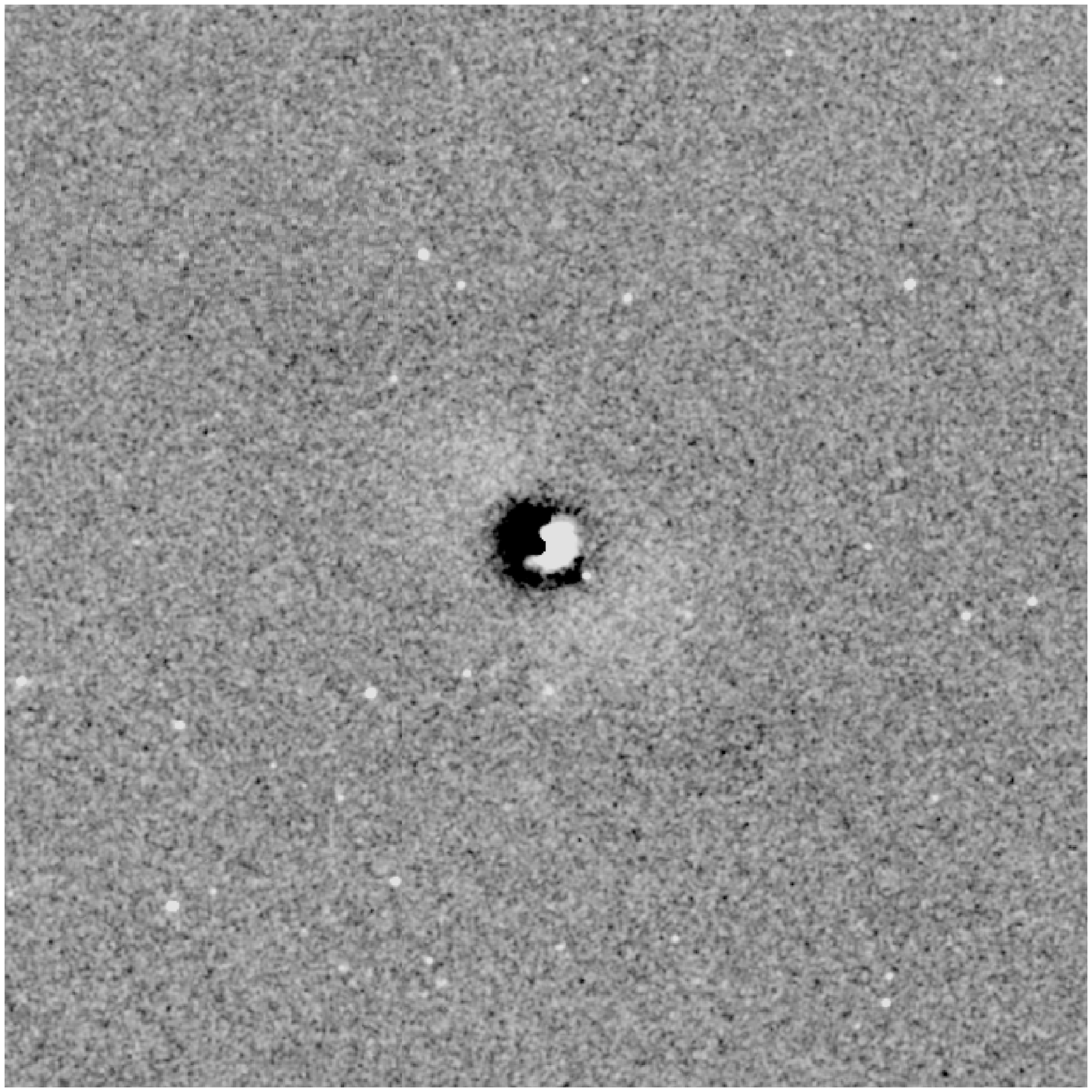}\\
\end{tabular}
\caption[GALFIT analysis of the S0 sample.]{GALFIT analysis of the S0
  sample. {\it Left panels}: galaxy image; {\it middle panels}: GALFIT
  model image; {\it right panels}: difference image, scaled to
  emphasize any residuals in the fit. The ratio between the flux in the residual image and in the scientific image is around $5\%$. From top to bottom: NGC~3115 in
  a $[800''\times800'']$ box, NGC~7457 in a $[600''\times600'']$ box,
  NGC~2768 in a $[800''\times800'']$ box, NGC~3489 in a $[\simeq
  300''\times300'']$ box and NGC~3384 in a $[\simeq 600''\times600'']$
  box. The dimension of the box corresponds to the dimension of the
  image in which the fit was performed, apart from NGC~3384, which was
  fitted simultaneously with the two companion galaxies appearing in
  the 2MASS $[1200''\times1200'']$ image, in order to remove any
  residual contamination. \label{fig:GALsample}}
\end{figure*}

The images were modeled using GALFIT \citep{Peng} to fit an
exponential disc and a Sersic profile spheroid.  For most galaxies, a
satisfactory fit was obtained by fixing the Sersic profile index at
the conventional de Vaucouleurs law value of $n=4$, but in the case of
NGC~2768 this produced a poor fit, so the index was left as a free
parameter.  The resulting fits to the images are
presented in Figure~\ref{fig:GALsample}, and the associated values for
the best fit parameters are given in Table~\ref{tab:galfit}.
As is apparent from Figure~\ref{fig:GALsample}, in some cases this
simple two component model fits the galaxy very well, while in others
systematic residuals indicate that the system is somewhat more
complex.  In particular, we find:
\begin{itemize}
\item NGC~3115 contains an additional very thin disc-like structure as
  well as the thicker disc component that we have fitted.  In
  addition, the de Vaucouleurs fit is not perfect for the bulge at
  very small radii.  
\item NGC~3489 contains a faint but significant
  ring structure.
\item NGC~3384 shows a central dipolar structure in the residuals,
  seemingly indicative of an off-centre nucleus.  
\end{itemize}
Although all these features are interesting, and tell us that even the
plainest looking S0 galaxy can be quite complex, they are all either
localized in regions of high surface brightness where we do not detect
the PNe used in this kinematic analysis, or they are of low surface
brightness compared to the main bulge and disc components. 
Accordingly, they do not compromise our ability to use the simple two
component fit to determine the relative contributions of disc and
spheroid to the light at each point in the galaxy.  

\begin{figure*}
\begin{tabular} {lr}
\includegraphics[width=0.4\textwidth]{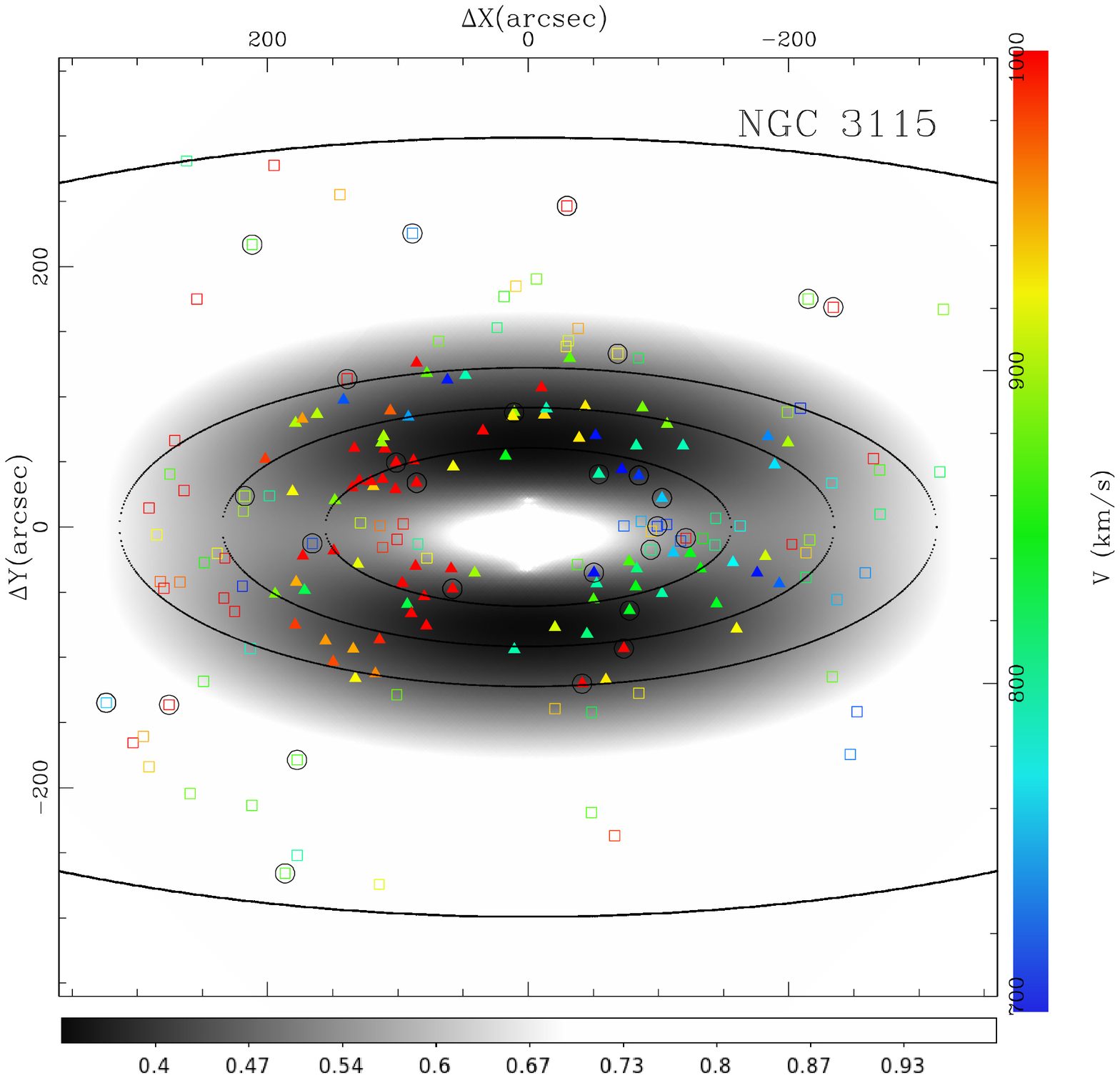} &  \includegraphics[width=0.4\textwidth]{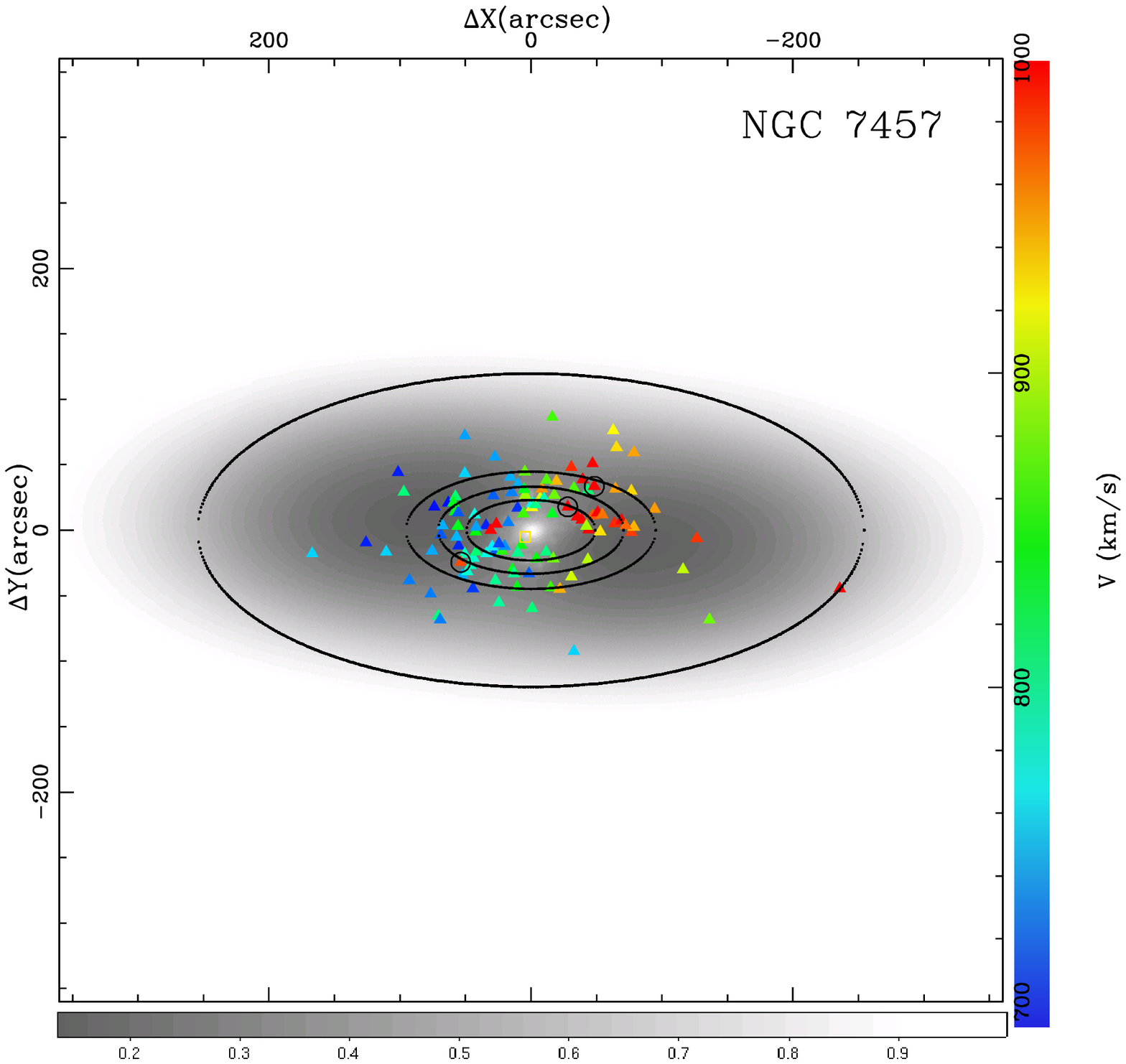}\\
\includegraphics[width=0.4\textwidth]{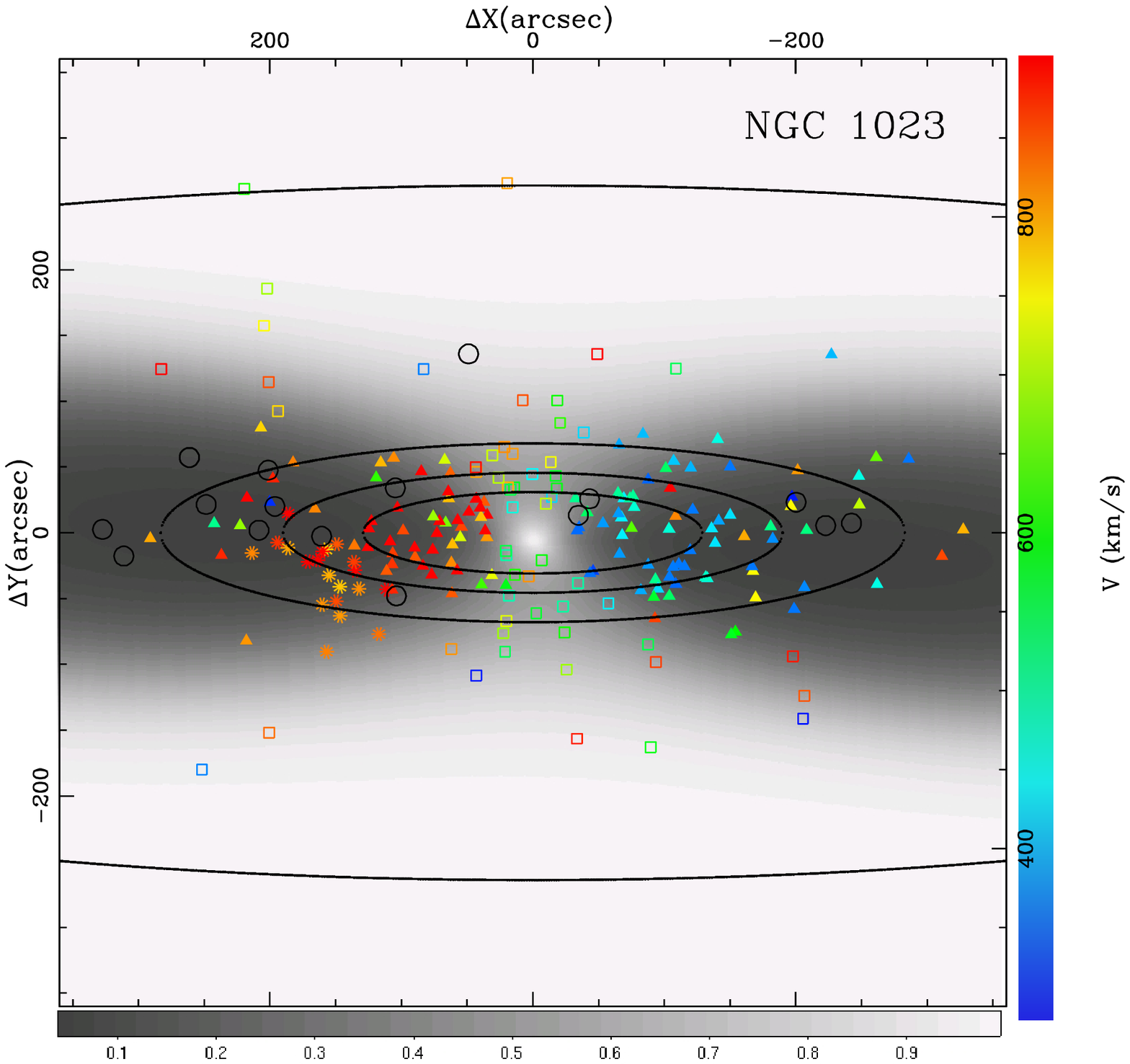}&\includegraphics[width=0.4\textwidth]{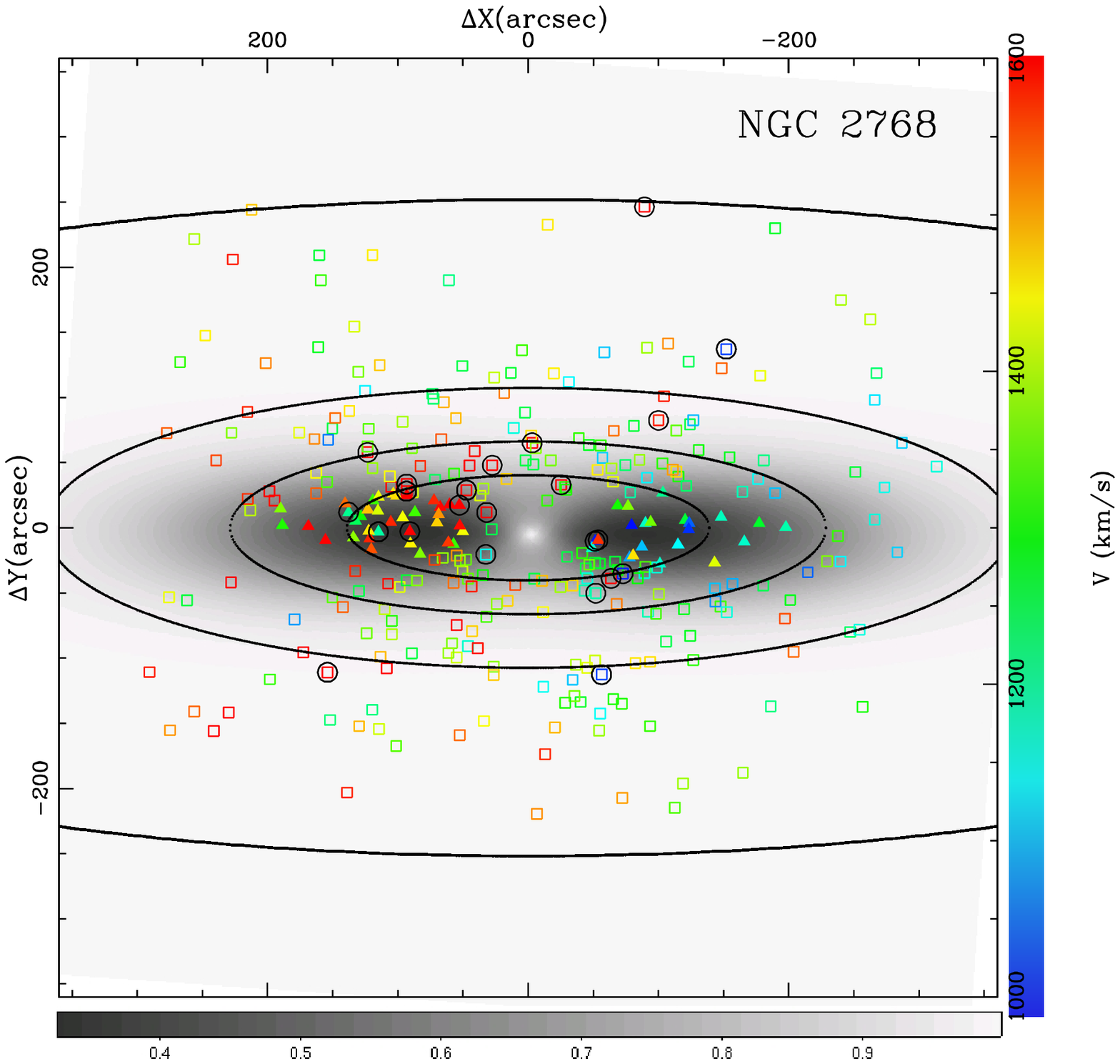}\\
\includegraphics[width=0.4\textwidth]{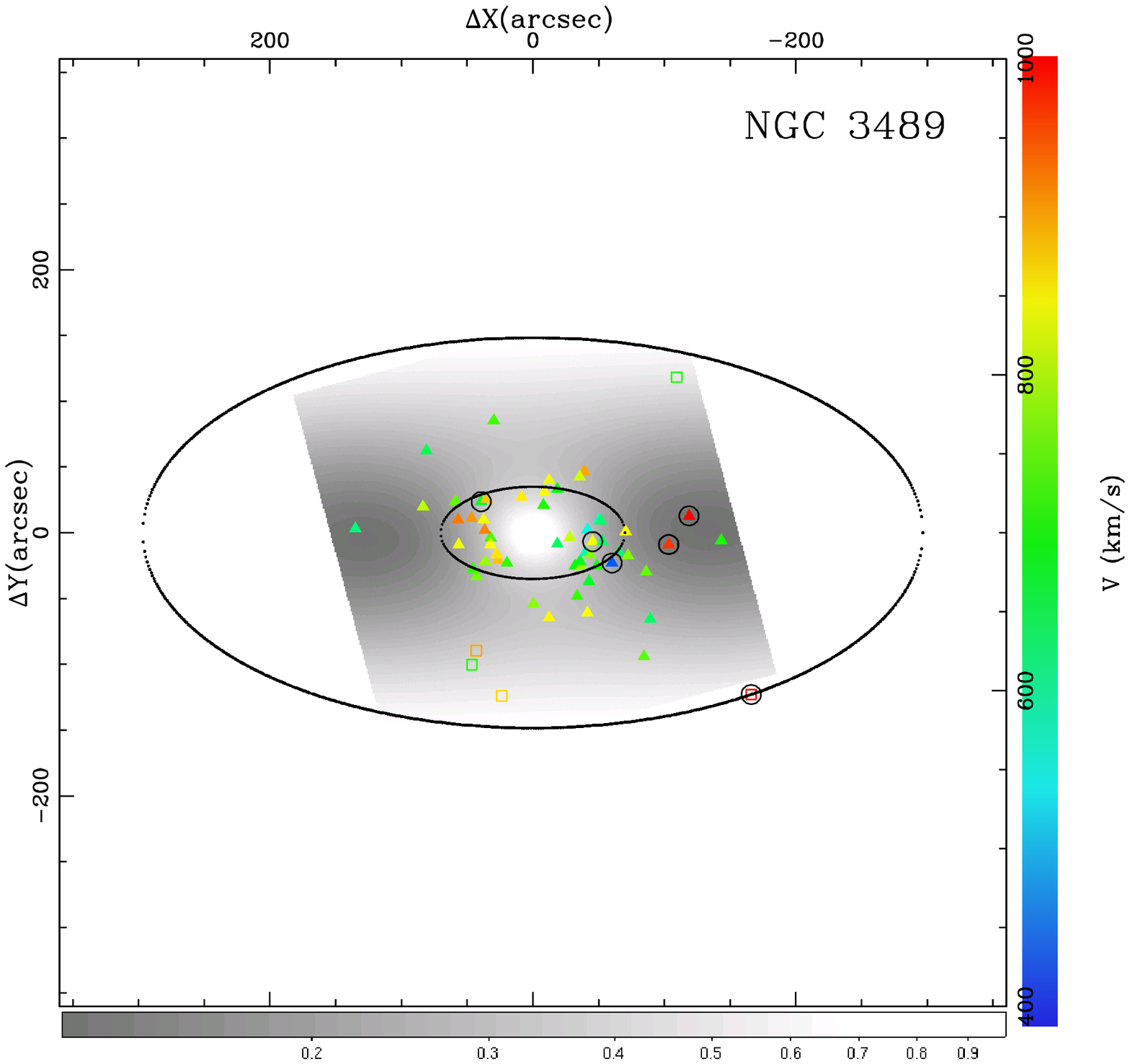} &\includegraphics[width=0.4\textwidth]{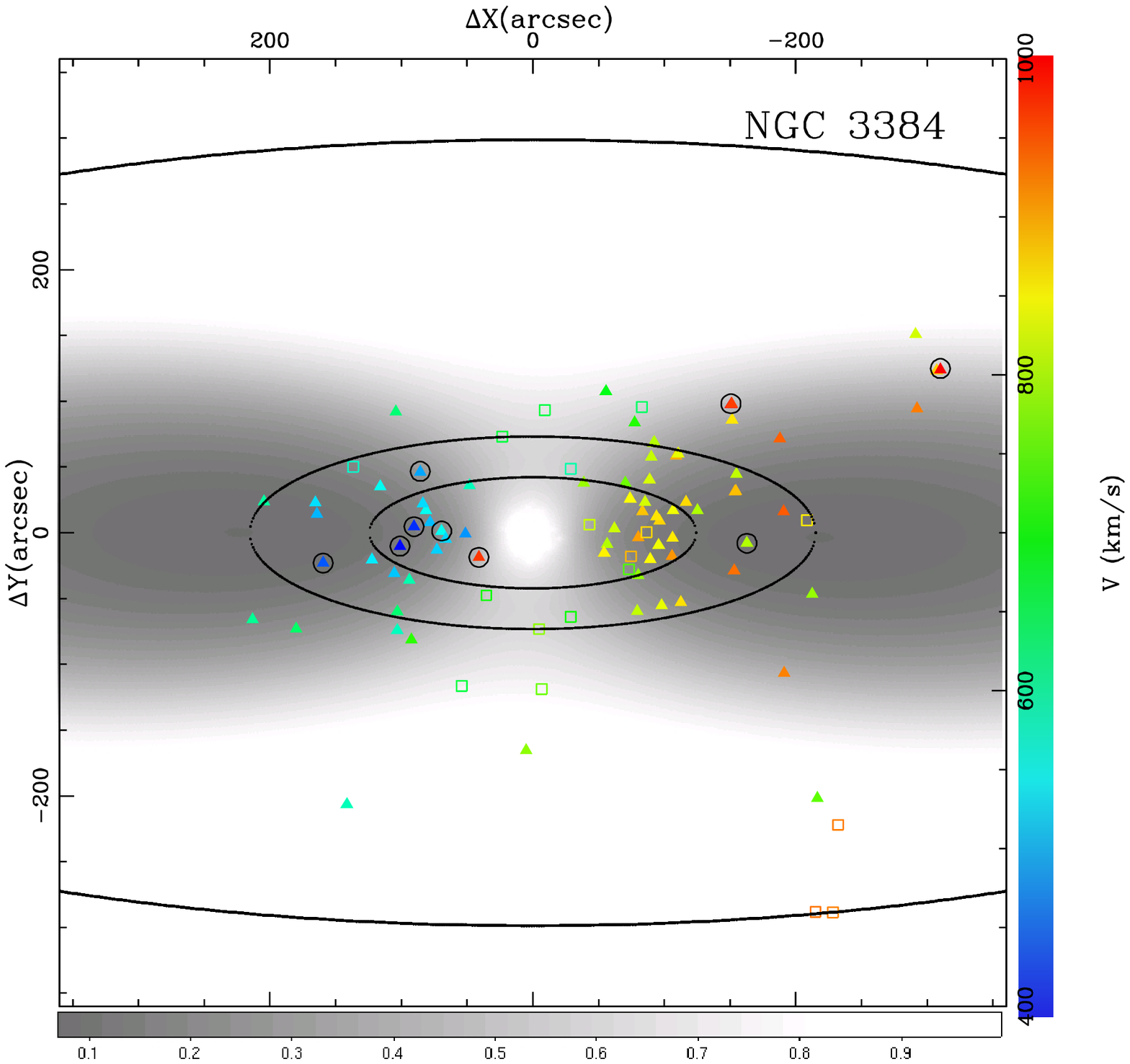}\\
\end{tabular}
\caption{Probability map that a PN belongs to the disc (darker areas)
  or spheroid (lighter areas).  The locations of detected PNe are also
  shown, with those more likely to belong to the disc shown as
  triangles, and those more likely to belong to the spheroid as
  squares.  Colours indicate the line-of-sight velocities of the PNe.
  Open circles are drawn around PNe rejected by the likelihood fit to
  the kinematics.\label{fig:sampleover}}
\end{figure*}

To illustrate the assignment of PNe to the two components,
Figure~\ref{fig:sampleover} shows a greyscale image of the
spheroid-to-total light at each point in the model.  The PNe are also
plotted and identified by whether they lie in a region where spheroid
or disc dominates, but note that, in the kinematic likelihood analysis
below, each is assigned an exact value $f_i$ from this image, such
that it has a probability $f_i$ that it comes from the spheroidal
component and $1 - f_i$ that it comes from the disc component
\citep{Ari}.

\section{disc and spheroid kinematic profiles}
\label{sec:disc and spheroid kinematics}

Having obtained the disc--spheroid light decomposition for each
galaxy, we can now fit all the line-of-sight velocities for the PNe
shown in Figure~\ref{fig:sampleover} using a maximum likelihood fit.
The method, described in detail in \citet{Ari}, essentially involves
fitting in radial bins using a model comprising a simple Gaussian
spheroid line-of-sight velocity distribution plus a similar Gaussian
velocity distribution for the disc component.  For the disc component,
we allow a non-zero mean velocity to fit rotation, varying with
azimuth as geometrically required for a rotating disc, and also
incorporate the fact that its line-of-sight velocity dispersion is a
different projection of the disc's coupled radial and tangential
velocity dispersions, $\sigma_{r}$ and $\sigma_{\phi}$ respectively,
at different azimuths.  We neglect the contribution of the
$z$-component of the velocity dispersion: it is intrinsically smaller,
and its modest projection along the line of sight further reduces its
significance in these inclined systems.  We also assume that the disc
is not too hot, so we can invoke the epicycle approximation to couple
the values of $\sigma_{r}$ and $\sigma_{\phi}$. This fitting process
thus solves for a simplified model of both disc and spheroid
kinematics, and also allows us to identify and reject  PNe that do
not fit the model, based on their individual contributions to the
total likelihood.

\begin{figure*}
\begin{tabular} {lr}
\includegraphics[width=0.4\textwidth]{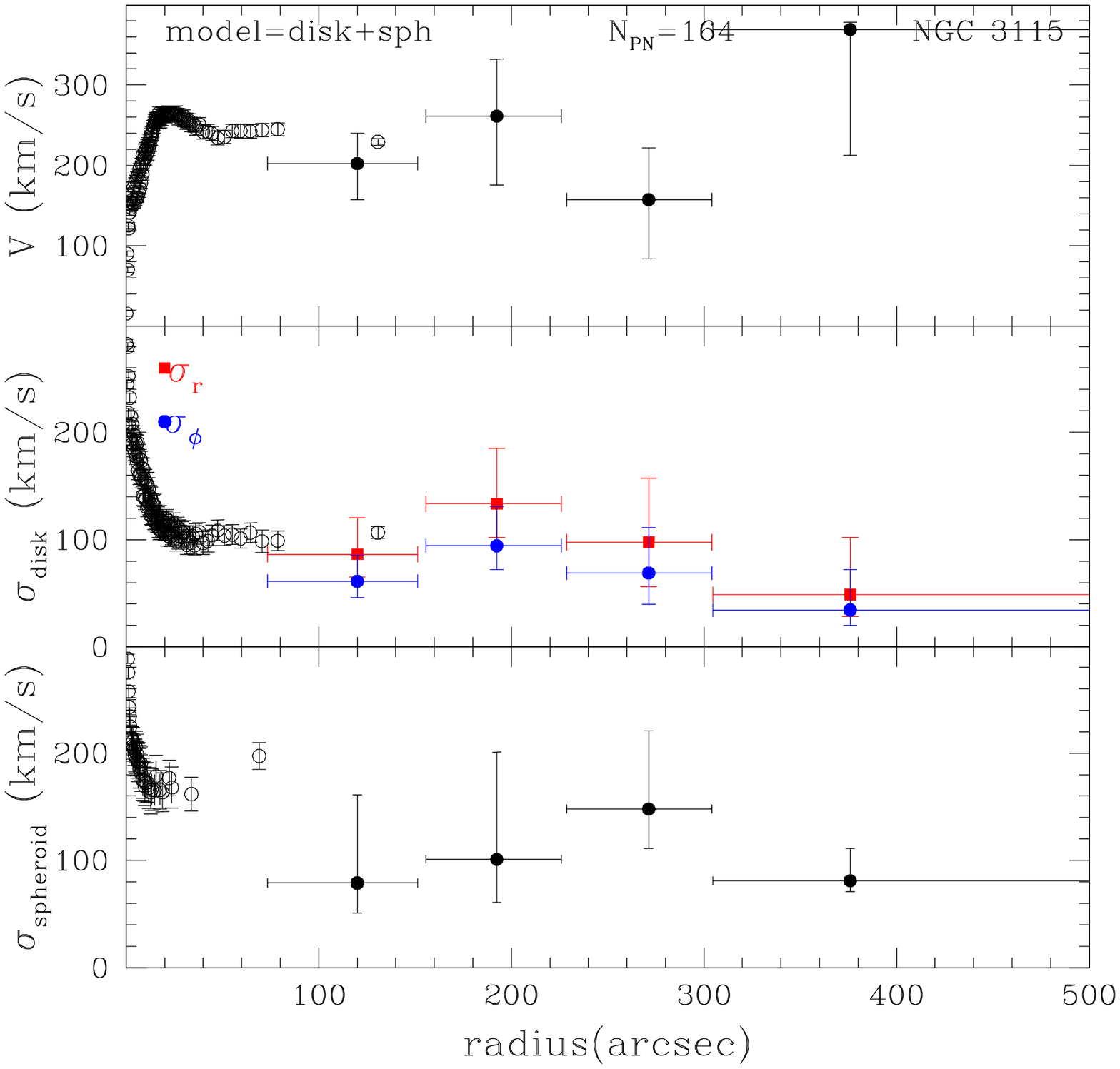} &  \includegraphics[width=0.4\textwidth]{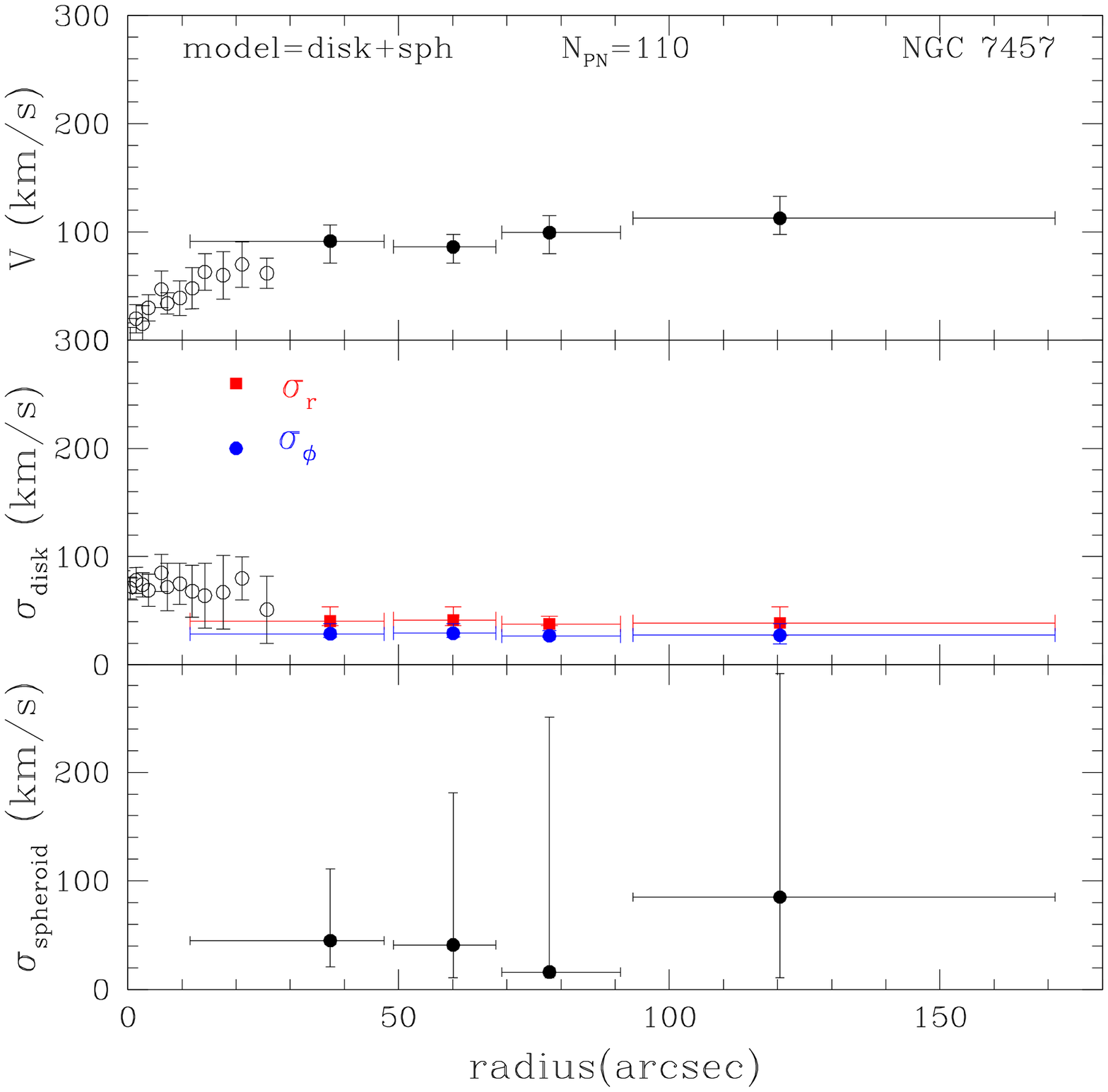}\\
\includegraphics[width=0.4\textwidth]{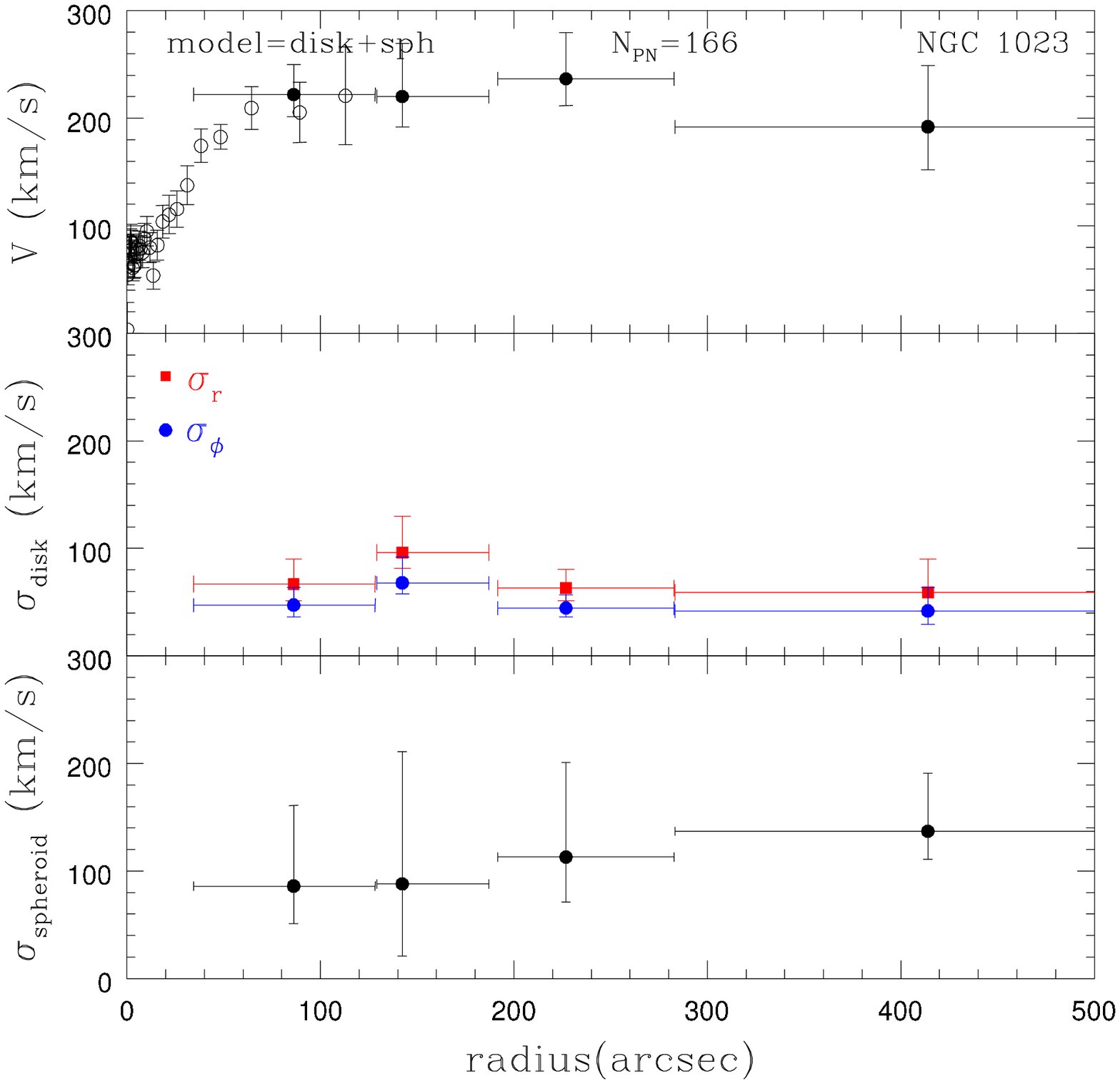}&\includegraphics[width=0.4\textwidth]{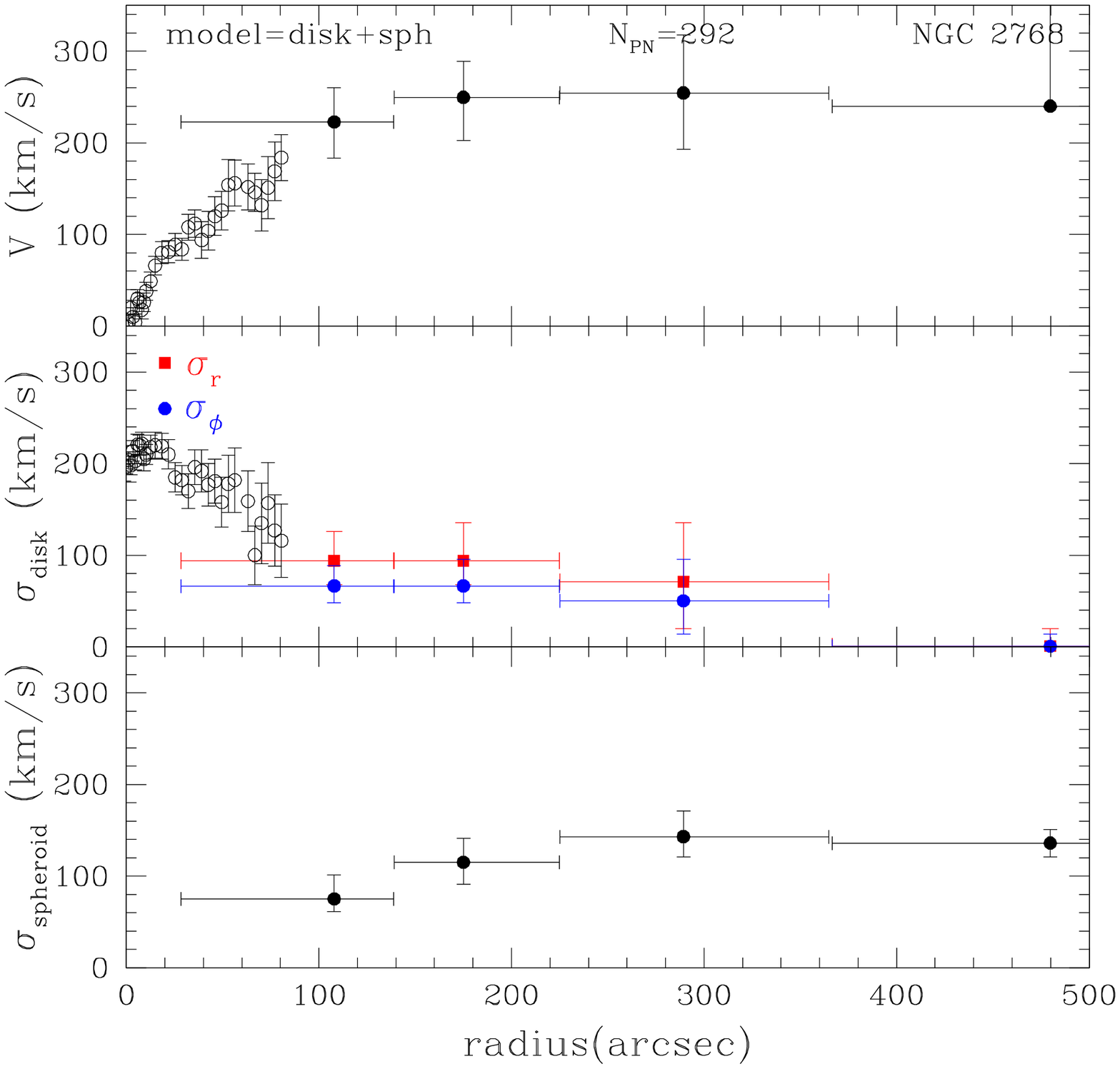}\\
\includegraphics[width=0.4\textwidth]{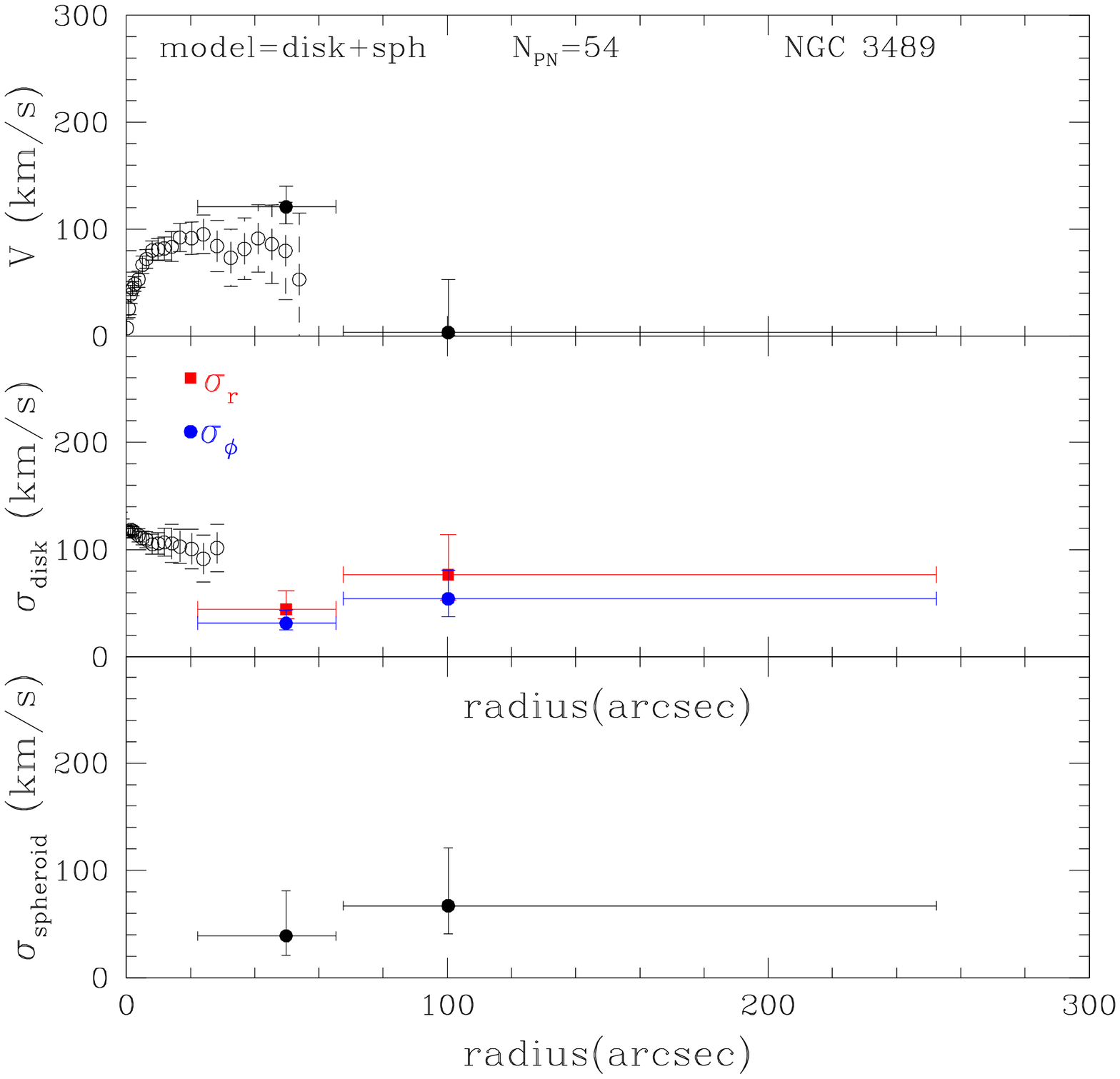}&\includegraphics[width=0.4\textwidth]{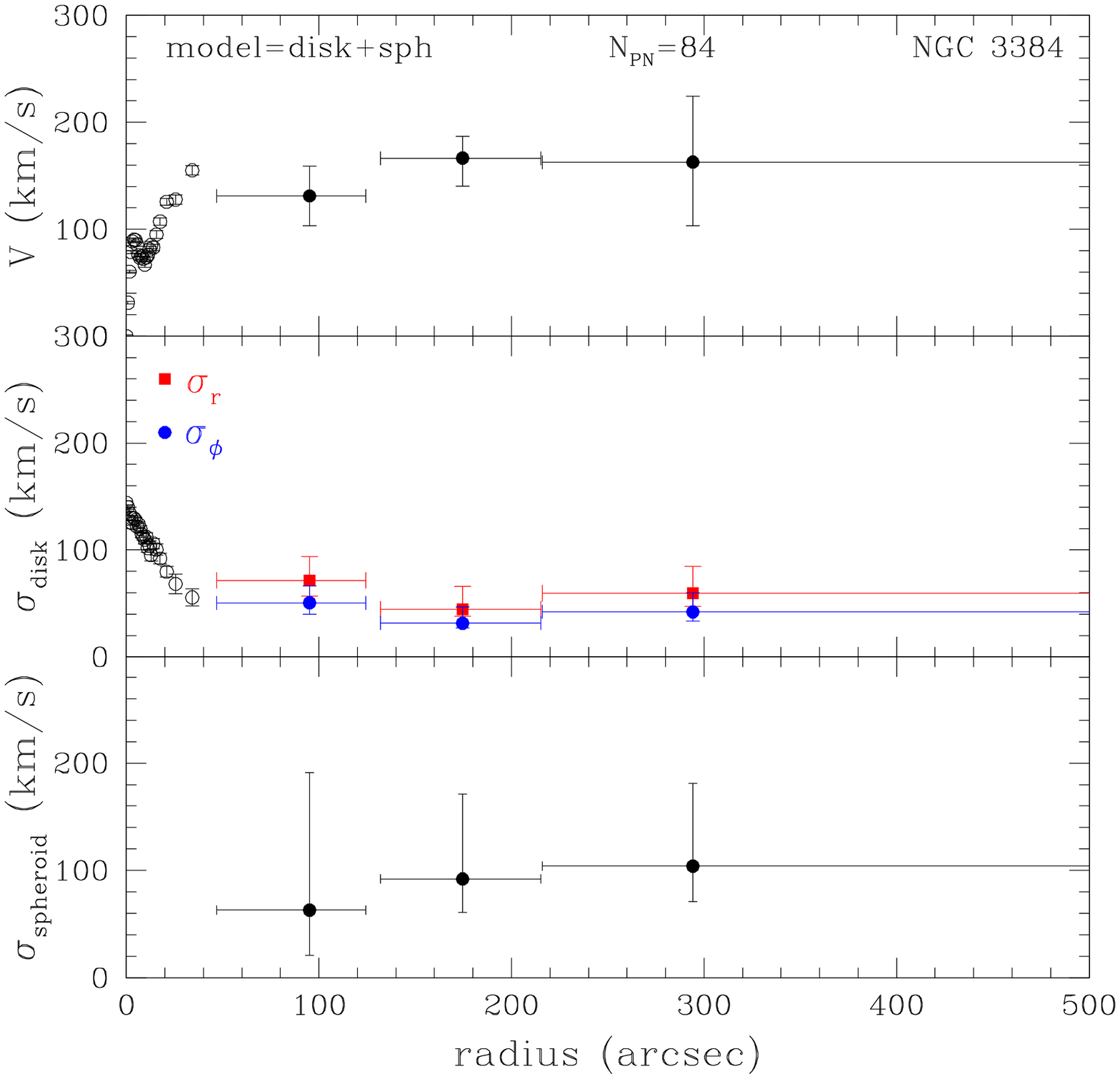}\\
\end{tabular}
\caption{disc and spheroid kinematics of the sample galaxies assuming a spheroid$+$disc
  model.  For each galaxy, we show the rotation velocity of the disc
  (top panel), its coupled tangential, $\sigma_{\phi}$, and radial, $\sigma_{r}$ dispersions (middle
  panel), and the velocity dispersion of the spheroid (bottom
  panel).  Vertical error bars indicate uncertainty, while horizontal
  error bars show the radial binning.  Where available, we also show
  kinematics derived from conventional absorption-line
  spectra (open circles) \citep{Caon,Simien,Norris,Deb}. \label{fig:sampleds}}
\end{figure*}
  
Any PNe rejected in this process are highlighted in
Figure~\ref{fig:sampleover}, and the final resulting best-fit
kinematic parameters as a function of radius are shown in
Figure~\ref{fig:sampleds}.  In each case, the radial bins have been
chosen to ensure that each contains at least 30 PNe, as found to be a
suitable minimum in \citet{Ari}, which is why the bin sizes and number
vary from galaxy to galaxy.  The resulting elliptical bin boundaries
are shown in Figure~\ref{fig:sampleover}.  For comparison,
Figure~\ref{fig:sampleds} also shows conventional absorption-line
kinematics along the major axes from various sources
\citep{Caon,Simien,Norris,Deb}.  Since the disc light dominates in
this region, we compare these data to the derived disc kinematics.  In
one case there is also published minor-axis data \citep{Norris} where
the bulge light dominates, so the absorption-line dispersion profile
is compared to the bulge kinematics that we derive.  In all cases, the
agreement between the two methods is good, but the comparison
underlines how conventional absorption-line spectroscopy is typically
limited to the bright inner parts of each galaxy.

One simplification that we have made in this analysis is in assuming
that the spheroidal component is non-rotating.  In general, there is
not sufficient data to allow us to relax this assumption, but in the
case of NGC~2768 \citet{Forbes2012} found that the larger number of
PNe allowed this extra degree of freedom to be introduced.  However,
the resulting spheroidal component was found to be completely
dominated by random motions, so the difference was negligible,
justifying the assumption in the case of this galaxy and rendering it
plausible for the other galaxies where there were not sufficient data
to test it directly. This assumption also fits with the findings in
spiral galaxies, where classical bulges tend to be very slowly
rotating \citep{MacArthur2009} while pseudo-bulges with low Sersic
indices display more rotational support \citep{Fabricius}; the high
Sersic indices for these galaxies (see Table~\ref{tab:galfit})
suggests we might expect a corresponding lack of rotation.  

One further assumption we made is that the galaxies have only two
components, a disc and a spheroid. As discussed above, the images show
some departures from this simple model, although in most cases these
extra features are fairly minor.  The only exception is NGC~3115,
which seems to contain two discs of different thickness.  This object
would clearly be interesting to study in its own right with a more
sophisticated model, but for the current analysis, it makes more sense
to treat it in a manner consistent with the other galaxies, fitting
only the spheroid and the brighter disc.  It is interesting to note
that this model results in more PNe being rejected from the fit than
in other galaxies, which once again illustrates the power of the
likelihood approach to reject discrepant objects and home in on the
components being fitted, as was originally seen in \citet{Ari}.

As is apparent from Figure~\ref{fig:sampleds}, this analysis reveals a
remarkably consistent kinematic picture amongst these galaxies.  The
one exception seems to be NGC~3489, although it is notable that this
galaxy contains the smallest number of available kinematic tracers,
which may be compromising the results somewhat.  However, apart from
this galaxy, all the systems here display kinematics with many
features in common: the spheroidal components have a dispersion that
varies very little with radius, and the disc components rise to flat
rotation curves with ordered motions dominating over random velocities
at all radii.  It is interesting to compare this situation with the
much more heterogeneous picture presented in Figure~\ref{fig:lambda};
it would appear that at least some of the variety in that plot arose
from the net effect of superimposing multiple kinematic components
whose respective contributions vary from galaxy to galaxy.  This new
consistency provides some confidence in the analysis, but also offers
at least the possibility that these galaxies are sufficiently
homogeneous for some common underlying formation mechanism to exist.

\section{Analyzing the characteristic kinematics}
\label{sec:study of the recovered kinematics}

\begin{table*}
\centering
\begin{tabular}{c|cccc}
\hline
Name & $(V_{*})_{\rm disc}$ & $(V_{*}/\sigma_{\phi})_{\rm disc}$  & $(V_{c})_{\rm disc}$ & $\hat{\sigma}_{sph}$\\ 
$$       & [km/s]             &    $$            &  [km/s]&           [km/s]                                  \\
\hline
\hline
NGC~3115  & $220_{-49}^{+41}$ &  $3.3_{-0.7}^{+0.6}$ & $264_{-61}^{+67}$ & $107_{-15}^{+36}$ \\
NGC~7457  & $113_{-22}^{+18}$ &  $3.7_{-0.7}^{+0.6}$ & $136_{-20}^{+23}$ & $54_{-20}^{+85}$ \\
NGC~2768  & $232_{-41}^{+39}$ &  $3.3_{-0.6}^{+0.6}$ & $316_{-70}^{+79}$ & $123_{-9}^{+11}$ \\
NGC~1023  & $244_{-33}^{+37}$ &  $5.3_{-0.7}^{+0.8}$ & $274_{-41}^{+47}$ & $112_{-18}^{+42}$ \\
NGC~3489  & $144_{-23}^{+37}$ &  $4.2_{-0.7}^{+0.5}$ & $171_{-27}^{+38}$ & $57_{-17}^{+36}$ \\
NGC~3384  & $179_{-28}^{+22}$ &  $5.3_{-0.8}^{+0.6}$ & $196_{-30}^{+37}$ & $89_{-19}^{+53}$ \\
\hline
\end{tabular}
\caption[Kinematic results]{Kinematic results. From left to right: galaxy name [1], rotation velocity in the disc at $3R_{d}$ [2], ratio between rotation and velocity dispersion along the tangential direction for the galaxy disc calculated at $3R_{d}$ [3], circular speed in the disc at $3R_{d}$ [4], light weighted spheroid dispersion velocity [4]. \label{tab:kineres}}  
\end{table*}

\subsection{Deriving characteristic values}\label{sec:charkin}
Having determined the kinematic profiles of the spheroid and disc
components of these S0 galaxies, we can now start to use these data to
seek archaeological evidence as to how they formed.  As a starting
point, we translate these profiles into characteristic values for the
kinematics of each component.  For the disc, we determine the mean
streaming motion, $V_*$, and the two components of velocity
dispersion, $\sigma_\phi$ and $\sigma_r$, at three disc scale lengths
as determined by the photometric parameters (see
Table~\ref{tab:galfit}), since by this radius they seem to have
settled to their asymptotic values.  The only exception is NGC~3489,
where the limited amount of kinematic data restricts us to calculating
these quantities at $2R_d$; in practice this makes little difference,
as in the other galaxies the parameters do not change significantly
over this radial range.  The resulting values are presented in
Table~\ref{tab:kineres}.  In the spheroids, the
velocity dispersion does not change significantly with radius, so we
simply calculate a luminosity-weighted average value,
\begin{equation}
\hat{\sigma}^{2}_{sph}=\frac{\sum_{i} L_{sph, i}\sigma^{2}_{sph,i}}{\sum_{i} L_{sph, i}},
\end{equation}
where $L_{sph, i}$ is the luminosity of the spheroid in each radial
bin, as ascertained from the photometric fit.  The resulting values of
$\hat{\sigma}_{sph}$ are also listed in Table~\ref{tab:kineres}. 

The other kinematic quantity we need to derive is the characteristic
circular speed of each galaxy, $V_c$, which provides a measure of the
system's mass.  Because of the presence of significant random motions,
we cannot simply use the mean streaming speed of the stars, but must
correct these motions for asymmetric drift via the equation
\begin{equation}
V_{c}^{2}=V_{\star}^{2}+\sigma_{\phi}^{2}-\sigma_{r}^{2}(1+\frac{d \ln \nu}{d \ln r}+\frac{d \ln \sigma_{r}^{2}} {d \ln {r}})\label{eq:jeans}
\end{equation}
\citep{Binney2}.  If the random motions are not too large and the
rotation curve is close to flat, then the epicycle approximation
implies that $\sigma_{r} = \sqrt{2} \sigma_{\phi}$.  If we further use
the photometrically-derived exponential disc profile (see
Section~\ref{sec:decomposition}) to determine $\nu(r)$ and fit the
observed variation in velocity dispersion with radius using a further
exponential with its own scalelength,
\begin{equation}
\sigma_{r}^{2}=\sigma_{r}^{2}(0) \exp (-\frac{r}{R_{2}}),
\end{equation} 
the asymmetric drift equation simplifies to 
\begin{equation}
V_{c}^{2}(r)=V_{\star}^{2}+\sigma_{\phi}^{2}(-\frac{1}{2}+\frac{ r}{ R_{d}}+\frac{r} {R_{2}}),
\label{eq:cv general}
\end{equation} 
We can hence estimate $V_c$ at any radius using this equation;
Table~\ref{tab:kineres} lists the derived characteristic value of this
quantity for each galaxy at $3R_{d}$, which matches the fiducial radius used
for the other kinematic parameters; it is also at large enough radii
that the circular speed will have converged to its characteristic
asymptotic value.  A check on the validity of this simplified equation
is provided by comparing our results to those of \citet{Timothy}, who
carried out full anisotropic Jeans modelling \citep{Cappellari} on two
of the sample galaxies.  They obtained circular velocities of
$310\,{\rm km}\,{\rm s}^{-1}$ for NGC~2768 and $160\,{\rm km}\,{\rm
  s}^{-1}$ for NGC~3489, in good agreement with the values derived
here.

 \begin{figure}
\includegraphics[width=0.4\textwidth]{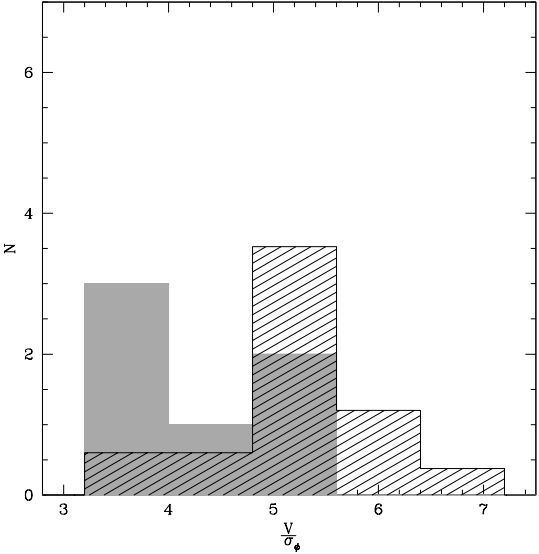}\\
\caption{Ratio of stellar rotation velocity in the disc to the disc
  velocity dispersion for the S0 galaxies of this sample (filled),
  and the corresponding quantity for a comparison sample of spiral
  galaxies from the literature (diagonally shaded)  \citep{Bottema,Herrmann2009}, normalised to the same number as the S0 galaxies. \label{fig:Bottema}}
\end{figure}

\subsection{Disc kinematics}
As previously mentioned, one key diagnostic of the evolutionary past
of a disc is provided by the ratio between its ordered and random
motions: if they are simply spiral galaxies that have ceased forming
stars, then they should be dominated by mean streaming motions in the
same way as their progenitors, whereas any more violent transition
such as one precipitated by a merger would tend to heat the disc and
hence decrease the dominance of the rotation.  Table~\ref{tab:kineres}
lists the characteristic values of $V_{*}/\sigma_{\phi}$ for these S0
systems, yielding a mean value of 4.2, with an RMS scatter of 0.8.
These values contrast with the $V_{*}/\sigma_{\phi} \sim 1$ obtained
by \citet{Bournaud} in the simulations that formed S0s from a range of
minor mergers, suggesting that this is not the mechanism responsible
for the creation of these galaxies.

To determine whether the data are consistent with the alternative
hypothesis that these discs are simply ``dead'' spiral discs, we
consider the kinematics of eight spiral galaxies presented by
\citet{Bottema} and five shown by \citet{Herrmann2009}, which together
form a comparison sample with a range of rotation velocities similar
to the S0s presented here.  These comparison data are unfortunately
somewhat heterogeneous in nature.  The \citet{Bottema} kinematic
values were derived at one disc scale length; since the velocity
dispersion tends to fall with radius while rotation increases, this
smaller fiducial radius will tend to produce smaller values of
$V_{*}/\sigma_{\phi}$ than at our adopted radius of $3R_d$.  The
comparison is more direct with the data from \citet{Herrmann2009}, as
we can interpolate a value at $3R_d$ from the kinematic profiles they
present.  In both cases, we again use the epicycle approximation to
interchange between $\sigma_r$ and $\sigma_\phi$.
Figure~\ref{fig:Bottema} shows the resulting estimates of
$V_{*}/\sigma_{\phi}$ for both the spiral comparison sample and the
current sample of S0s.  Although there is a reasonable degree of
overlap, implying that these S0s could have formed passively from
spirals, there are also strong indications that the values are
systematically lower for the S0s, which is not the sense one would
expect if it arose from the possible bias in the smaller fiducial
radii used by \citet{Bottema}.  Thus, if these S0s formed from a
random selection of comparable present-day spirals, it would seem that
some heating of their discs must have occurred during the transition.

Although the sample becomes small when divided in this way, there is
no evidence that the degree of heating of the disc depends on
environments, since the S0s from isolated, small group and large group
surroundings (as described in Section~\ref{sec:introduction}) do not
display systematically different values of $V_{*}/\sigma_{\phi}$.  We
are thus left with a developing picture in which the formation of an
S0 from a spiral appears somewhat more violent than a simple shutdown
of star formation, but not as extreme as a merger-induced transition,
with no indication that the mechanism depends strongly on environment.

\subsection{Scaling relations}
\label{sec:TFFJ}
We can seek further insight into the possible scenario for
transformation, and also try to address the question of whether S0s
are more closely related to spirals or ellipticals, by looking at
where these systems lie on the scaling relations respected by the
other galaxy classes.  

\begin{figure}
\includegraphics[width=0.45\textwidth]{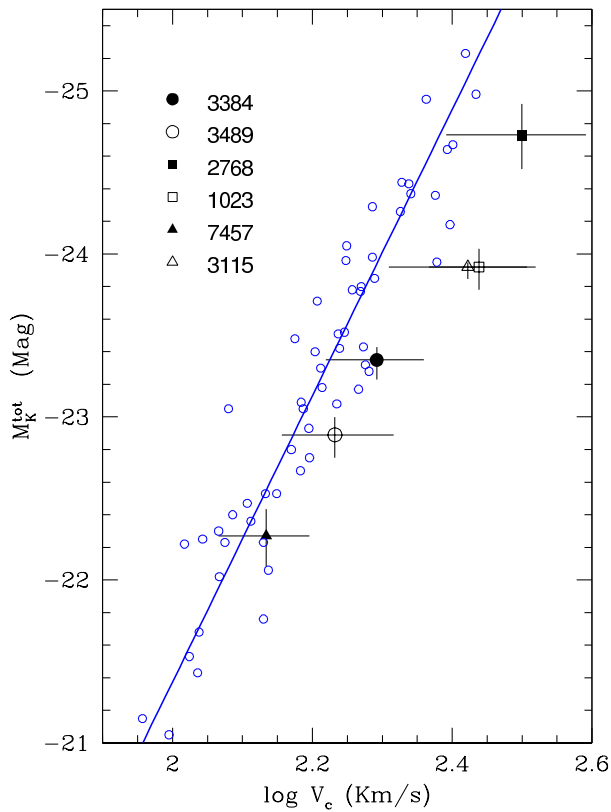}
\caption{$K$-band Tully--Fisher relation for the S0 galaxies analyzed
  here, compared to the relation for spiral galaxies (blue circles and
  line of best fit) from \citep{Tully2} and
  \citet{Rothberg2000}\label{fig:tully}}
\end{figure}

\subsubsection{Tully--Fisher relation}

The simplest test we can carry out is to place these S0 galaxies on
the Tully--Fisher relation, which displays a very tight correlation
between a spiral galaxy's luminosity and its circular rotation speed
\citep{Tully}.  Such comparisons have previously been carried out by
various authors, with somewhat mixed results \citep{Salamanca,
  Neistein, Williams, Pahre}.  One reason for the discrepant results
is the difficulty in obtaining the circular rotation speed: for the
spiral galaxies, this quantity can be reliably estimated from gas
kinematics, but for the S0s we are reliant on stellar motions.  The
previous studies using conventional absorption-line spectroscopy have
therefore faced two significant problems.  First, unlike PN data, the
absorption-line spectra do not typically reach particularly large
radii, so probe regions where the rotation curve has not reached its
asymptotic value and where the random motions are still large so the
somewhat-uncertain asymmetric drift correction is also large (see
Section~\ref{sec:charkin}).  Second, as we saw in
Section~\ref{sec:introduction}, the composite nature of the kinematics in
these multi-component systems can significantly affect the inferred
rotational properties, and it is only by separating the distinct
kinematic components as we have done here that we can expect to get a
reliable measure of the disc rotation speed and hence the galaxy's
circular speed.  We are therefore in a good position to place S0
galaxies reliably on the Tully--Fisher relation.  

The other element we need for these S0 galaxies are measurements of
their absolute magnitudes.  We carry out this analysis using the
$K$-band photometry introduced in Section~\ref{sec:decomposition} to
minimise the impact of dust obscuration.  There is a risk that the
relatively shallow 2MASS exposures might miss some of the flux from
the outer parts of the galaxy, so we use the magnitudes derived from
the full GALFIT models obtained in Section~\ref{sec:decomposition}
rather than the values quoted in the published 2MASS catalogue; in
fact, this effect turns out to be rather small, with an average offset
of only 0.05 magnitudes. For NGC~3489, we had to
use an SDSS $z$-band image to carry out the GALFIT modeling, so for
this galaxy we calculate a total $z$-band magnitude from the model,
then use the SDSS/2MASS integrated colours to convert it to a $K$-band
magnitude, which implicitly assumes that there are no strong colour
gradients in this galaxy.

To convert these values to absolute magnitudes, we use the distance
moduli derived from surface brightness fluctuations from
\citet{Tonry}, systematically decreased by $0.16$ mag to take into
account the updated Cepheid zero point of \citet{Freedman}, as
discussed in \citet{Coccato} and \citet{Ari2}.  Here, it is important
to use a distance scale that has not itself been based on
Tully--Fisher or other kinematic scaling relations, to avoid
introducing a circularity into the analysis.

Figure~\ref{fig:tully} shows the resulting Tully--Fisher relation for
the S0 galaxies analyzed here, compared to the relation for spiral
galaxies \citep{Tully2, Rothberg2000}, calibrated to the same absolute
magnitude scale.  As found previously by \citet{Salamanca}, there is
clearly an offset between the S0 and spiral relations, in the sense
that the S0s are systematically fainter.  This offset fits
straightforwardly with a picture in which the S0 galaxies ceased
forming stars at some point in the past, leading to a steady fading of
their stellar populations.  Again, there is no indication of any
variation in this offset with environment, suggesting that this is not
a driving factor in the timing of the transition, but there does seem
to be a trend in the sense that the most massive galaxies display the
largest offset; in the context of the time since evolving off the
spiral Tully--Fisher relation, this trend can be interpreted as
another instance of ``downsizing,'' in which more massive galaxies go
through evolutionary transitions at an earlier stage than less massive
galaxies.

\begin{figure}
\includegraphics[width=0.45\textwidth]{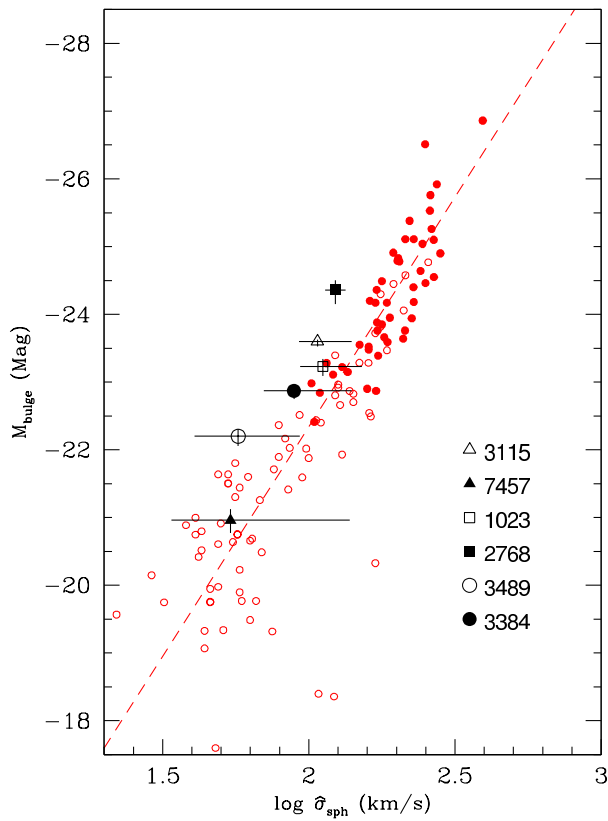}\\
\caption{$K$-band Faber--Jackson relation plotting spheroid absolute
  magnitude as a function of its dispersion for the S0 galaxies
  studied here.  Comparable kinematic data for elliptical galaxies
  from \citet{Mobasher99} (filled red circles) and \citet{Matkovic05}
  (open red circles) are also shown.  The line shows a fit to the
  brighter elliptical galaxy data.\label{fig:faber}}
\end{figure}

\subsubsection{Faber--Jackson relation}
Having separated spheroid and disc components, we are also in a
position to look at connections in the other direction along the
Hubble sequence to see how closely S0s might be related to elliptical
galaxies.  For these systems, the equivalent scaling is the
Faber--Jackson relation between the spheroidal component velocity
dispersion and its absolute magnitude.

Accordingly we have plotted these quantities for the S0 galaxy
spheroids in Figure~\ref{fig:faber}, using the spheroid dispersions
calculated in Section~\ref{sec:charkin}, and spheroid absolute
magnitudes derived as described above.  For comparison, we have also
plotted the relation followed by elliptical galaxies as derived by
\citet{Mobasher99}.  These latter data are based on velocity
dispersions measured at an effective radius, away from any central
spike in velocity dispersion, so should be directly comparable to the
dispersions determined for the S0 spheroids.  These elliptical data do
not span the full range of magnitudes seen for the S0 spheroids, so we
supplement them with kinematic data on fainter early-type galaxies in
the Coma Cluster \citep{Matkovic05} using $K$-band magnitudes from
\citet{Skrutskie+06}.  For these fainter galaxies, the kinematics were
derived from fibre spectra, but the large three-arcsecond fibres used
will cover most of these small faint galaxies at the distance of Coma,
so should again be comparable to the luminosity-weighted average
values of $\hat{\sigma}_{sph}$ derived for the S0 spheroids. To obtain
the absolute magnitude for the Coma galaxies, we placed the
apparent magnitudes tabulated in \citet{Mobasher99} and
\citet{Skrutskie+06} at a distance modulus of $m-M = 35.06$, as also
found using surface brightness fluctuations \citep{Thomsen97}.  The
fact that the two comparison data sets follow a common relation even
though their kinematics were derived with very different radial
weightings indicates that the results are unlikely to be sensitive to
the exact manner in which the average velocity dispersion is
obtained.  

The spheroids of the S0s follow a trend similar to the
Faber-Jackson relation for ellipticals, although they seem to lie
along its upper envelope.  One possible explanation for such an offset
might be that the influence of the disc surrounding the S0 spheroid might in some
way decrease its velocity dispersion.  However, simulations by
\citet{Debattista} show that the formation of a disc around
such a spheroidal component would in fact serve to compress it and
increase its velocity dispersion, so the presence of a disc does not
explain the difference.  A simpler explanation is that the offset is
along the other axis, due to some enhancement of the spheroid light in
S0s when compared to other elliptical systems.  


\begin{figure}
\includegraphics[width=0.45\textwidth]{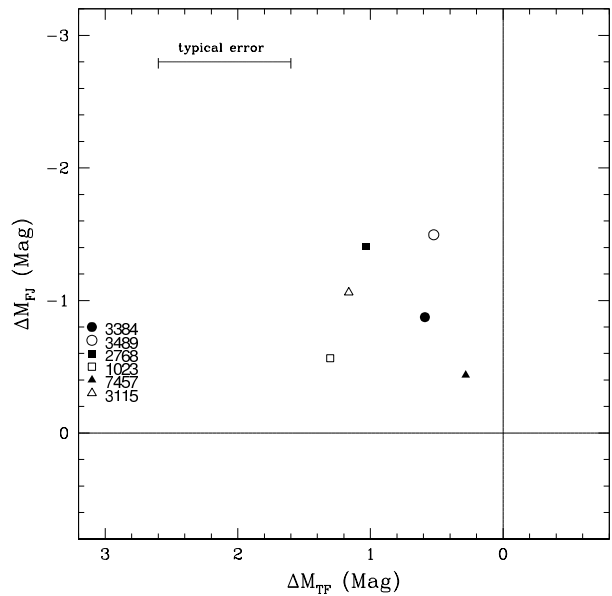}\\
\caption{Plot showing how the S0 galaxies are offset in magnitude from the mean
  Tully--Fisher and Faber--Jackson relations for spiral galaxies and
  elliptical galaxies respectively.\label{fig:TFFJ}}
\end{figure}

\section{Discussion}
\label{sec:Discussion}

In this paper, we have analysed the kinematics of a sample of six
lenticular galaxies using the planetary nebulae data presented in
\citet{Ari2} to determine their dynamical properties out to large radii.
This analysis has emphasised the importance of treating the spheroidal
and disc components of these systems as distinct entities both
photometrically and kinematically, and doing a careful job of
separating them.  Through this process, the rather heterogeneous
properties of S0s over-all start to be resolved into more consistent
sub-components, and a reasonably coherent picture begins to emerge.

The common factors we have uncovered are that:
\begin{enumerate}
\item The discs of S0 galaxies are comparable to those of spirals,
  with similar flat rotation curves and falling velocity dispersion
  profiles, but with a larger amount of random motions.
\item The spheroids of S0 galaxies show flat dispersion profiles,
  similar to what is found in some ellipticals.
\item S0 galaxies follow the Tully--Fisher relation, but offset to
  fainter magnitudes than spiral galaxies, with a greater offset for
  more massive galaxies.
\item S0 galaxy spheroids follow the Faber--Jackson relation, somewhat 
  offset to brighter magnitudes than elliptical galaxies.
\item There is no strong evidence that any of these effects depend on
  the current environment of the galaxy.
\end{enumerate}
The Tully--Fisher and Faber--Jackson findings are summarised in
Figure~\ref{fig:TFFJ}, which shows the offset from each relation for
the S0 sample.  In the absence of any systematic effect, we would
expect these points to appear equally in all four quadrants, but in
fact they appear strongly clustered in the region of the plot where
the spheroids are too bright but the overall galaxies are too faint.

With this level of detail starting to become available, we can begin
to sketch out a plausible evolutionary sequence leading to the
formation of S0 galaxies.  If these systems began their lives as
spiral galaxies, at some point they underwent a transition that
removed their gas supply and hence cut off star formation.  The
process responsible does not depend strongly on current environment,
but does seem to have acted earlier on more massive galaxies as
another example of downsizing.  The hotter discs of S0s could indicate
that their high-redshift spiral progenitors had less ordered discs
\citep{Kassin}, and that this larger amount of random motion became
locked in when star formation ceased.  However, if the transformation
occurred at a point when these systems had settled down to today's cold
spiral systems, the hotter discs would then indicate that the galaxies
had undergone a process sufficiently disruptive to heat the system
significantly, but less violent than a merger.  

This aggressive process might well also dump some fraction of the gas
being stripped from the disc into the centre of the galaxy, causing a
last burst of turbulent star formation there, enhancing the spheroid's
luminosity significantly, shifting it off the Faber--Jackson relation.
Such a scenario would also fit with the recent finding that the
spheroids of S0 galaxies have systematically younger stellar
populations than their discs \citep{Johnston12}.

Of the various mechanisms advanced for causing galaxy transformation,
perhaps the most plausible to realize such a scenario is a form of
the ``harassment'' invoked by \citet{Moore96} as a process for
transforming galaxies in clusters.  In their paper, they demonstrated
that this process of repeated high-speed encounters could strip gas
from a galaxy while dumping some at the centre of the system under
transformation as we require here.  However, we also know that the
process needs to be somewhat gentler than they modelled because we
need to preserve the disc structure of the galaxy with only a fairly
modest amount of heating.  It remains to be seen whether such mild
harassment, perhaps better thought of as ``pestering,'' could be made
to work across the range of environments in which S0s are found, but
we do at least now have a growing number of observational constraints
against which any such model can be tested.

\section*{Acknowledgements}
AC acknowledges the support from both ESO (during her studentship in
2011 and the visitor program in 2012) and MPE (visitor program 2012).
LC acknowledges funding from the European Community Seventh Framework
Programme (FP7/2007-2013/) under grant agreement No 229517. AJR was
supported by National Science Foundation grant AST-0909237. We thank the referee for the supportive and constructive report. The PN.S
team thanks both the UK and NL time allocation committees and the
staff of the WHT for their strong support in acquiring the data used
in this research. This research has also made use of the 2MASS data
archive, the NASA/IPAC Extragalactic Database (NED), and of the ESO
Science Archive Facility, for which we are grateful.


\bibliography{refs}
\bibliographystyle{mnras_biblist}

\end{document}